\documentclass[hidelinks,11pt]{article}

\newcommand{\ner}{\boldsymbol{r}}

\newcommand{\de}{\,\mathrm{d}}                               
\newcommand{\e}{\operatorname{e}}                               
                     
\newcommand{\inc}{\mathrm{inc}}

\newcommand{\andtext}{\quad\mbox{and}\quad}

\newcommand{\p}{\partial}

\newcommand{\real}{\mathrm{Re}\,}                                   
\newcommand{\imag}{\mathrm{Im}\,}

\newcommand{\lf}{\left}
\newcommand{\rg}{\right}

\newcommand{\R}{\mathbb{R}}       
\newcommand{\C}{\mathbb{C}}

\newcommand{\mgf}{\bold H}                                        
\newcommand{\elf}{\bold E}

\newcommand{\bnor}{\bold n}          
\usepackage{multirow}
\usepackage{amsmath}
\usepackage{amsfonts}
\usepackage{amsmath}
\usepackage{amssymb}  
\usepackage{graphicx}
\usepackage{caption}
\usepackage{mathrsfs}
\usepackage{upgreek}
\usepackage{subfig}
\usepackage{booktabs}
\usepackage{authblk}
\usepackage{stmaryrd}

\newcommand{\rjmp}{\rg\rrbracket}
\newcommand{\ljmp}{\lf\llbracket}
\newcommand{\lavg}{\left\{{\hskip -6pt}\left\{}
\newcommand{\ravg}{\right\}{\hskip -6pt}\right\}}
\newcommand{\slavg}{\left\{{\hskip -4.2pt}\left\{}
\newcommand{\sravg}{\right\}{\hskip -4.2pt}\right\}}

\newcommand{\circint}{\mathop{\mathpalette\docircint\relax}\!\int}
\newcommand{\docircint}[2]{%
  \ifx#1\displaystyle
    \displaycircint
  \else
    \normalcircint{#1}%
  \fi
}
\newcommand{\displaycircint}{\displaystyle\mathsf{c}\mkern-25mu}
\newcommand{\normalcircint}[1]{%
  \smallerc{#1}\ifx#1\textstyle\mkern-9mu\else\mkern-8.2mu\fi
}
\newcommand{\smallerc}[1]{%
  \vcenter{\hbox{$\ifx#1\textstyle\scriptstyle\else\scriptscriptstyle\fi\mathsf{c}$}}%
}

\title{Sideways adiabaticity: Beyond ray optics for slowly~varying metasurfaces}
\author[1,2]{Carlos P\'erez-Arancibia\thanks{cperezar@mit.edu}}
\author[1,3]{Rapha\"el Pestourie}
\author[1]{Steven~G.~Johnson}
\affil[1]{\small{Department of Mathematics, Massachusetts Institute of Technology, Cambridge, MA 02139, USA}}
\affil[2]{\small{Institute for Mathematical and Computational Engineering, School of Engineering and Faculty of Mathematics, Pontificia Universidad Cat\'olica de Chile, Santiago, Chile}}
\affil[3]{\small{School of Engineering and Applied Sciences, Harvard University, Cambridge, MA 02138, USA}}
\date{\today}

\begin{document}
\maketitle

\begin{abstract}
Optical metasurfaces (subwavelength-patterned surfaces typically described by variable effective surface impedances) are typically modeled by an approximation akin to ray optics: the reflection or transmission of an incident wave at each point of the surface is computed as if the surface were ``locally uniform'', and then the total field is obtained by summing all of these local scattered fields via a Huygens principle. (Similar approximations are found in scalar diffraction theory and in ray optics for curved surfaces.) In this paper, we develop a precise theory of such approximations for variable-impedance surfaces. Not only do we obtain a type of \emph{adiabatic theorem} showing that the ``zeroth-order'' locally uniform approximation converges in the limit as the surface varies more and more slowly, including a way to quantify the rate of convergence, but we also obtain an infinite series of \emph{higher-order corrections}.  These corrections, which can be computed to any desired order by performing integral operations on the surface fields, allow rapidly varying surfaces to be modeled with arbitrary accuracy, and also allow one to validate designs based on the zeroth-order approximation (which is often surprisingly accurate) without resorting to expensive brute-force Maxwell solvers. We show that our formulation works arbitrarily close to the surface, and can even compute coupling to guided modes, whereas in the far-field limit our zeroth-order result simplifies to an expression similar to what has been used by other authors.
\end{abstract}

\section{Introduction}\label{sec:intro}

Optical metasurfaces, subwavelength structures described by an effective sheet impedance~\cite{kuester2003averaged,Holloway:2012bv,holloway2009discussion,holloway2011characterizing,Tretyakov:2015cr,Tcvetkova:2018jf,epstein2014passive,Epstein:2014bx,Pfeiffer:2013by,achouri2015general}, are now being designed for large-area optical devices using models in which the far-field reflection/transmission coefficients are computed at each point assuming a \emph{uniform} (or periodic) surface---as explained below, we refer to these as ``ray-optics'' models.  This is a good approximation for surfaces (or unit cells) that are varying slowly, a fact that is closely connected to the ``adiabatic theorem''~\cite{katsenelenbaum1998theory,johnson2002adiabatic} for waves propagating through slowly varying media.  However, although there are countless papers and books on modeling propagation \emph{through} slowly varying media~\cite{marcuse1974theory,snyder2012optical,katsenelenbaum1998theory}, exploiting the rate of change as a small parameter $\varepsilon$, the ``sideways'' problem of scattering \emph{off} a slowly varying surface (Fig.~\ref{fig:config}) is relatively unstudied. In this paper, we address the following key questions: how quickly does the ray-optics approximation converge as $\varepsilon \to 0$, can we quickly compute the low-order corrections (both to improve accuracy and to validate ray optics), and how do we compute both far-field and near-field scattering (e.g. coupling to guided modes)? {A typical metasurface has two scales: the subwavelength scale of the microstructure and the macroscale of the nonuniformity. In this paper, we address corrections due to the macroscale nonuniformity, which we allow to be completely arbitrary, while we follow other authors~\cite{kuester2003averaged,Holloway:2012bv,holloway2009discussion,holloway2011characterizing,Tretyakov:2015cr,Tcvetkova:2018jf,epstein2014passive,Epstein:2014bx,Pfeiffer:2013by,achouri2015general} in subsuming the microscale into effective surface impedances.}

In particular, we use the technical machinery of surface-integral equations (SIEs)~\cite{colton2013integral,nedelec2001acoustic} and a ``locally uniform approximation" of the metasurface to show that the ray-optics approximation is the far-field zero-th order term in a convergent series~(Section~\ref{eq:Somm_int} and Appendix~\ref{sec:integral_eq_corrections}), that each successive correction can be computed simply {by} performing integrals (not by solving any PDE or other system of equations), and that the next-order correction scales as~$\varepsilon^2$. Moreover, our series allows us to compute the full Green's function of the surface: the fields in response to arbitrary sources or incident fields, including the near-field terms (fields and/or sources close to the surface). We show that these near fields allow us to compute the coupling of an incident wave to guided modes on the surface~\cite{martini2014metasurface,Tcvetkova:2018jf} and that they also appear in the zeroth-order locally uniform approximation.  For rapidly varying metasurfaces, such as those designed to reflect light at a very oblique angle~\cite{wong2016reflectionless}, we show that even the far-field accuracy is substantially improved by including the next-order correction. Perhaps more importantly, the ability to compute the next-order correction provides a way to \emph{validate} a ray-optics design for very large-area metasurfaces, where brute-force Maxwell simulations are impractical and there was previously no way to evaluate the ray-optics accuracy short of a laboratory experiment.

Since typical metasurface designs lead to large computational domains (often hundreds of wavelengths~\cite{Khorasaninejad:2017gg}) that are intractable by standard simulation techniques, e.g. finite-difference and finite-element methods, previous work on metasurfaces has made extensive use of numerical simulations based on ray-optics approximations. In particular, authors typically compute reflection/transmission coefficients for \emph{periodic} surfaces with a variety of unit cells, they assume that these coefficients remain accurate even for an aperiodic surface where each unit cell is different, and then they select the unit cell at each point on the surface to achieve a desired optical functionality~\cite{verslegers2010phase,Pfeiffer:2014bv,achouri2015general,Aieta:2012gm,yu2013flat,Yu:2011eya,Yu:2014hq,raphael_paper}.   (For subwavelength unit cells where there is only a single ``specular'' reflected/transmitted wave, the reflection/transmission coefficients can also be fitted to an effective sheet impedance, giving a ``homogenized'' effective medium at each point~\cite{niemi2013synthesis,selvanayagam2014polarization,pfeiffer2014bianisotropic,achouri2015general}.)    Because these works described the surfaces by a single far-field (planewave) reflection and/or transmission coefficient at each point, they can be thought of as  ``ray-optics'' approximations even if they were expressed in the language of wave optics.  (A closely related approximation---curved surfaces treated as locally flat---is called a ``tangent-plane'' or ``Kirchhoff'' approximation~\cite{voronovich2013wave}. Yet another closely related approximation is provided by scalar diffraction theory~\cite{o2004diffractive}.) Here, we assume that the metasurface is subwavelength enough to be described as an effective sheet impedance at each point, but we do \emph{not} only compute the scattering assuming that the impedance is locally uniform: our goal is to take the \emph{macro}-scale spatial variation (the aperiodicity) explicitly into account by computing correction to the locally uniform approximation.  (Potential extensions to slowly varying periodic structures where the micro-scale is treated explicitly, perhaps to include additional diffraction orders for large-period structures, are discussed in Section~\ref{sec:conclusions}.)

Wave propagation \emph{through} slowly varying media is usually treated by coupled-mode theory: one expands the wave in the basis of ``instantaneous''~\cite{cohen1973quantum} eigenfunctions of each cross-section~\cite{marcuse1974theory} or period~\cite{johnson2002adiabatic}, and then obtains a set of coupled differential equations in the mode coefficients (typically truncated to only a few guided modes).  As the medium varies more slowly, the coefficients tend to constants, corresponding to ``adiabatic'' transport of modes without inter-modal scattering~\cite{marcuse1974theory}.  Unfortunately, this approach appears awkward to apply to the problem of scattering \emph{off} of a slowly varying medium, both because there is a \emph{continuum} of radiating modes and because one wants to describe the basis of incoming/outgoing planewaves independently of the varying surface in order to connect to a ray-optics (zero-th order) approximation. The fact that an incident planewave is not an eigenfunction of the cross-section at each point means that one cannot simply quote the standard adiabatic theorem to justify the metasurface ray-optics approximation, for example.   Another type of technique for approximating the scattering from a weakly perturbed surfaces is a Born approximation~\cite{chew1995waves,snyder2012optical} (also known as a volume-current method~\cite{johnson2005roughness}, Kirchhoff approximation~\cite{voronovich2013wave}, etc.), which handles perturbations like surface roughness that are small in amplitude but \emph{not} necessarily slowly varying, whereas our goal is to handle variations that are slow but not small.

A surface impedance $Z(\varepsilon x)$ (defined precisely in Section~\ref{eq:prob_form}) varies more and more slowly as $\varepsilon\to 0$, and our goal is an expansion with terms proportional (in a certain norm) to powers of~$\varepsilon$~\cite{colton2013integral,nedelec2001acoustic}. This expansion is achieved through an SIE. In particular, since the media above and below the surface are homogeneous, we express the problem in terms of an SIE in which the unknowns reside only on the surface. Our derivations start with an approximate Green's function $G_{\rm p}$ that is a building-block of the locally uniform approach (Section~\ref{eq:Somm_int}), and then we insert this into an exact SIE, obtained by enforcing to transition conditions on the metasurface, to derive a series of corrections (Appendix~\ref{sec:integral_eq_corrections})---like a Born--Dyson series~\cite{born1999principles}, the corrections are expressed in terms of integrals involving $G_{\rm p}$. These integrals must be computed numerically by a ``quadrature'' technique~\cite{davis2007methods} (Appendix~\ref{sec:numerics}), but such computations are simple summations on a computer that are far easier than solving the large systems of equations arising in brute-force computational methods, and also have the advantage of parallelizing perfectly (fields at different points can be computed completely independently). {Truncating to the zeroth-order term in the series does \emph{not} correspond to setting $\varepsilon=0$ (a uniform surface), so even the lowest order locally uniform approximation captures to certain extent the surface variation.} In the far field, $G_{\rm p}$ simplifies to an expression $G^{\rm ff}_{\rm p}$  that can be written in closed form (eliminating an integral), recovering the usual ray-optics approximation at zero-th order (Appendix~\ref{sec:far_field}).   Using this approach, we demonstrate through numerical experiments that the ray-optics approximation (i.e., the far-field of the locally uniform approximation) produces far-field errors that vanish as~$\varepsilon^2$, and more generally as $\varepsilon^{2N+2}$ if we include $N$th-order corrections (Fig.~\ref{fig:ex_convergence}). In the presence of guided modes, which correspond to poles that appear in $G_{\rm p}$ at certain wavevectors~\cite{chew1995waves}, we show that this also simplifies the integrals in our perturbative expansion (via a steepest-descent approximation) if one is mainly interested in coupling to guided modes (Appendix~\ref{sec:far_field} and Figs.~\ref{fig:ex_beam} and~\ref{fig:ex_5}).

\section{Problem formulation}\label{eq:prob_form}
\begin{figure}[h!]
\centering	
\includegraphics[scale=0.9]{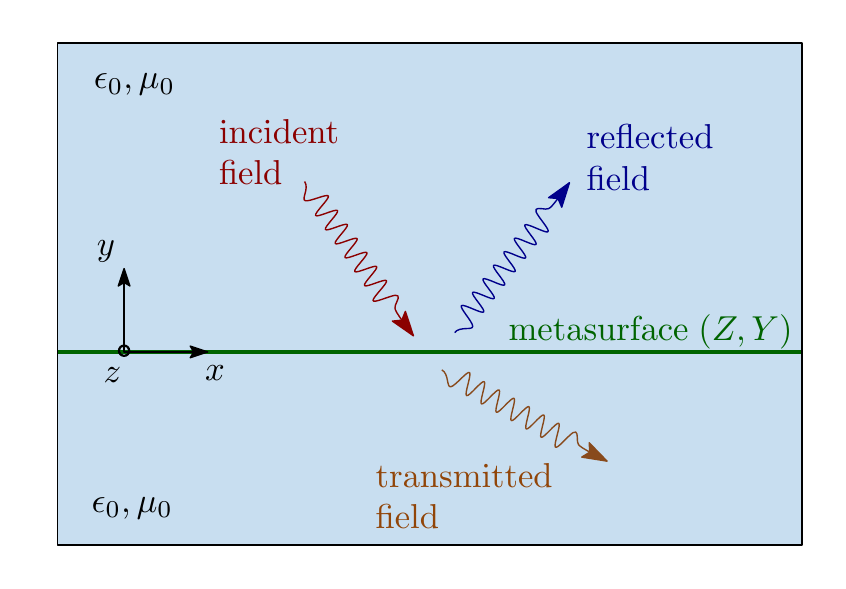}
\caption{Scattering problem under consideration. A time-harmonic incident field $u^\inc$ with angular frequency $\omega>0$ impinges on a  metasurface with surface impedance $Z$ and surface admittance $Y$,  producing a reflected field propagating in the upper half-plane and a transmitted field propagating in the lower half-plane. The metasurface is surrounded by a homogeneous medium with electric permittivity $\epsilon_0>0$ and magnetic permeability $\mu_0>0$. }\label{fig:config}
\end{figure}

We consider a metasurface  $\Gamma$ in two spatial dimensions  that divides the $xy$~plane into two unbounded  half-planes, $\Omega_+=\{y>0\}$ and $\Omega_-=\{y<0\}$, occupying the regions above and below $\Gamma$, respectively. The media $\Omega_+$ and $\Omega_-$ surrounding the metasurface are assumed to be homogeneous with electric permittivity and magnetic permeability denoted by $\epsilon_0>0$ and $\mu_0>0$, respectively (Fig.~\ref{fig:config}).  The metasurface is characterized by the so-called generalized sheet transition conditions~\cite{kuester2003averaged}:
\begin{equation}
\begin{split}
Z\bnor\times \lf(\mgf^{+}-\mgf^{-}\rg)=&\ \frac{1}{2}\lf(\elf^{+}_{\parallel}+\elf^{-}_{\parallel}\rg),\\
-Y\bnor\times\lf(\elf^{+}-\elf^{-}\rg)=&\ \frac{1}{2}\lf(\mgf^{+}_\parallel+\mgf^{-}_\parallel\rg),
\end{split}\label{eq:trans_cond}
\end{equation}
which for the sake of presentation simplicity are assumed to be given in terms of scalar quantities corresponding to surface impedance $Z$ and the surface admittance $Y$~\cite{epstein2014passive}.  The symbol $\bnor$ in~\eqref{eq:trans_cond} denotes the unit normal vector to $\Gamma=\{y=0\}$ pointing upwards, and $\elf^\pm_\parallel$ and $\mgf^\pm_\parallel$ (resp. $\elf^\pm$ and $\mgf^\pm$) denote the tangential (resp. entire) fields at $y=0^\pm$.

It thus follows from Maxwell's equations that in $E_z$~polarization the total electromagnetic field $(\elf,\mgf)$ is given by $\elf=(0,0,E)$ and $\mgf=(H_1,H_2,0)$, and can be obtained from the $z$-component of the electric field by means of  the relations:
\begin{equation}
\nabla^2 E+k^2 E = 0,\quad H_1=\frac{1}{ik\eta}\frac{\p E}{\p y},\quad H_2=-\frac{1}{ik\eta}\frac{\p E}{\p x}, \label{eq:TE_relations}
\end{equation}
where $k= \omega\sqrt{\epsilon_0\mu_0}$ is the wavenumber of the surrounding  media, $\omega>0$ is the angular frequency, and  $\eta = \sqrt{\mu_0/\epsilon_0}$ is the intrinsic free-space impedance. Similarly, in $H_z$~polarization it holds that $\elf=(E_1,E_2,0)$ and $\mgf=(0,0,H)$ where 
\begin{equation}
\nabla^2 H+k^2 H=0,\quad E_1=-\frac{\eta}{ik}\frac{\p H}{\p y},\quad E_2=\frac{\eta}{ik}\frac{\p H}{\p x}.\label{eq:TM_relations}
\end{equation}

Relations~\eqref{eq:TE_relations} and~\eqref{eq:TM_relations}, on the other hand,
 yield that the transition conditions~\eqref{eq:trans_cond} can be equivalently expressed as
\begin{subequations}\begin{equation}\begin{split}
\lf\llbracket\frac{\p E}{\p y}\rg\rrbracket=&\ -\frac{ik\eta}{2Z}\slavg E\sravg \andtext \lavg \frac{\p E}{\p y} \ravg=-2ik\eta Y\ljmp E\rjmp,
\end{split}\end{equation}
in $E_z$ polarization, and 
\begin{equation}\begin{split}
\ljmp\frac{\p H}{\p y}\rjmp= -\frac{ik}{2Y\eta}\slavg H\sravg\andtext \lavg \frac{\p H}{\p y}\ravg=-\frac{2ikZ}{\eta}\ljmp H\rjmp,
\end{split}\end{equation}\label{eq:TE_TM_GSTC}\end{subequations}
in $H_z$ polarization, where the notations
\begin{equation}\label{eq:jumps_sums}
\ljmp u\rjmp =u^+-u^-\andtext \{\!\{ u\}\!\} =u^++u^-,
\end{equation}
have been introduced to refer to the jump and the sum of a scalar field $u$ across $\Gamma$, where $u^+$ (resp. $u^-$) denotes the limit value of  $u$ on $\Gamma$ from  $\Omega_+$ (resp. $\Omega_-$).

In order to treat both $E_z$- and $H_z$-polarization cases, we define the \emph{metasurface parameters} $\alpha$ and $\beta$ as 
\begin{subequations}\label{eq:surf_param_def}\begin{equation}
\alpha= \frac{\eta }{2Z}\andtext \beta =2\eta Y
\end{equation}
in $E_z$~polarization, and  by 
\begin{equation} 
 \alpha= \frac{1}{2Y\eta}\andtext \beta =\frac{2Z}{\eta}
 \end{equation}\end{subequations}
in $H_z$~polarization.  Throughout this paper we assume  that  $\alpha$ and $\beta$ are continuous complex-valued functions that satisfy $\real\alpha\geq 0$ and $\real\beta\geq 0$, which correspond to assuming that both the surface impedance $Z$ and the surface admittance $Y$ are passive but not necessarily lossless. We will eventually consider these quantities to be slowly varying functions of the form $\alpha(x)=a(\varepsilon x)$ and $\beta(x)=b(\varepsilon x)$, where $\varepsilon>0$ is a small parameter.

Letting  $u^{\rm tot}$ denote either the total electric field $E$ in $E_z$~polarization or the total magnetic field $H$ in $H_z$~polarization, it follows from~\eqref{eq:TE_TM_GSTC},~\eqref{eq:jumps_sums} and~\eqref{eq:surf_param_def} that the transition conditions  can be equivalently expressed as
\begin{equation}\begin{split}
 \ljmp u^{\rm tot}_y\rjmp=-ik\alpha\slavg u^{\rm tot}\sravg \andtext \lavg u^{\rm tot}_y\ravg=-ik\beta \ljmp u^{\rm tot}\rjmp\quad\mbox{on}\quad\Gamma,
\end{split}\label{eq:transition_conditions}\end{equation}
in terms of the metasurface parameters $\alpha$ and $\beta$ introduced in~\eqref{eq:surf_param_def}, where we have used the notation $u_y=\p u/\p y$.

In this paper we consider the problem of scattering that arise when the metasurface is illuminated by a time-harmonic incident field $u^\inc$ which is assumed to satisfy the Helmholtz equation $\nabla^2 u^\inc+k^2u^\inc=0$ in $\Omega_+$ and $\Omega_-$ (Fig.~\ref{fig:config}). 
In order to properly formulate a scattering problem, we proceed to express the total field as $u^{\rm tot}=u^{\rm scat}+u^\inc$, where $u^{\rm scat}$ denotes the \emph{scattered  field} off of~$\Gamma$. Replacing $u^{\rm tot}=u^{\rm scat}+u^\inc$ in the Helmholtz equation and the transition conditions~\eqref{eq:transition_conditions} we  obtain  that  $u^{\rm scat}$ satisfies
\begin{subequations}\begin{eqnarray}
\nabla^2 u^{\rm scat} + k^2 u^{\rm scat}&=& 0\quad \mbox{in}\quad \Omega_+\cup\Omega_-,\label{eq:HEqn}\\
\ljmp u^{\rm scat}_y\rjmp&=&-ik\alpha\lavg u^{\rm scat}\ravg-2ik\alpha u^\inc\quad\mbox{on}\quad\Gamma,\label{eq:bc_1}\medskip\\
\lavg u^{\rm scat}_y\ravg&=& -ik\beta\ljmp u^{\rm scat}\rjmp- 2u_y^\inc\quad\mbox{on}\quad\Gamma.\label{eq:bc_2}
\end{eqnarray}
In order for~\eqref{eq:huygens_scattering} to be a well-posed boundary value problem for $u^{\rm scat}$, the scattered field has to satisfy a certain radiation condition at infinity~\cite{colton2013integral,bleistein2012mathematical,nosich1994radiation}. Such a radiation condition, which roughly speaking means that $u^{\rm scat}$ corresponds to an up-going wave-field in $\Omega_+$ and a down-going wave-field in $\Omega_-$, can be formally stated in terms of the angular spectral representation by requiring the existence of functions (or more generally, distributions) $A_+$ and $A_-$ such that 
\begin{equation}
u^{\rm scat}(\ner) = \int_{C} A_\pm (k_x) \e^{ik_x x+i\sqrt{k^2-k_x^2}|y|}\de k_x\quad\mbox{for}\quad \ner=(x,y)\in\Omega_\pm,
\end{equation}
where contour $C$ corresponds to the real $k_x$-axis that is suitably dented around the possible poles singularities of  $A_\pm$~\cite{DeSanto:1996vz}.\label{eq:huygens_scattering}\end{subequations}

\section{Exact and approximate surface integral representations}
In this section we derive an exact and an approximate integral representation formulae for the scattered field~$u^{\rm scat}$ solution of~\eqref{eq:huygens_scattering}. Such formulae involve the incident surface currents and the \emph{Green's function} of the boundary value problem~\eqref{eq:huygens_scattering} and are given in terms of integrals on the metasurface only, 

\subsection{Exact integral representation}\label{sec:exact_rep}
The Green's function $G$ of the boundary problem~\eqref{eq:huygens_scattering} can be physically interpreted  as the total field produced by a point source excitation placed off of the metasurface~\cite{chew1995waves,bleistein2012mathematical}. In detail, letting  $\ner' = (x',y')$ denote the location of a point source and $\ner=(x,y)$ denote an observation point, the Green's function $G(\ner|\ner')$ can be found by solving the following boundary value problem: 
{\begin{subequations}\begin{eqnarray}
\nabla^2_{\ner} G(\ner|\ner') + k^2 G(\ner|\ner')&=& -\delta(\ner-\ner'),\quad  \ner=(x,y)\in\Omega_+\cup\Omega_-,\label{eq:Helm_PS}\\
\ljmp G_y(\ner|\ner')\rjmp &=&-ik\alpha(x)\slavg G(\ner|\ner')\sravg,\quad \ner\in\Gamma,\label{eq:BC_alpha}\\
\lavg G_y(\ner|\ner')\ravg&=&-ik\beta(x)\ljmp G(\ner|\ner')\rjmp,\quad \ner\in\Gamma, \label{eq:BC_beta}
\end{eqnarray}\label{eq:huygens_Green}\end{subequations} }
with $\delta$ denoting the Dirac's delta distribution and where $G$ is additionally required to satisfy the radiation condition.

As is shown in Appendix~\ref{app:symmetry}, Green's third identity together with~\eqref{eq:huygens_scattering} and~\eqref{eq:huygens_Green} can be combined to show that the scattered field  $u^{\rm scat}$ admits the integral representation 
{\begin{equation}\label{eq:exact_rep_formula}
u^{\rm scat}(\ner) =\int\displaylimits_{-\infty}^\infty\lf\{G(\ner|s,0^+)f^\inc_+(s)-G(\ner|s,0^-)f^\inc_-(s)\rg\} \de s,\quad \ner\in\Omega_+\cup\Omega_-,\end{equation}}
where the current densities $f_\pm^\inc$ in~\eqref{eq:exact_rep_formula}  are  given by 
\begin{equation}
f^\inc_{\pm}(s) =u_y^\inc(s,0)\pm ik\alpha(s) u^\inc(s,0)\label{eq:RHS}
\end{equation}
in terms of the incident field $u^\inc$, and where $G(\ner|s,0^\pm) = \lim_{\delta\to 0^+} G(\ner|s,\pm\delta)$. 

The integral representation formula~\eqref{eq:exact_rep_formula} provides an explicit expression for the scattered field which is valid everywhere (in the near and far fields). It has, however, little practical relevance unless an exact or approximate Green's function is available. Unfortunately, a formula for the Green's function~\eqref{eq:huygens_Green} cannot be easily obtained for ``general" spatially varying metasurface parameters $\alpha$ and $\beta$, and thus, suitable approximations of $G$ are needed in order to make proper use of~\eqref{eq:exact_rep_formula} in scattering simulations. In the next section we derive an approximation for $G$ based on ray-optics principles.

\subsection{Ray-optics  approximation\label{sec:k_approx}}

From the viewpoint of Huygens' principle (formalized by the principal of equivalence), equation~\eqref{eq:exact_rep_formula} represents the scattered field by a source term corresponding to each point along the wavefront incident upon the surface ($f^\inc_\pm$)~\cite{born1999principles}. The typical "ray-optics" approximation  is to compute the reflection/transmission at each point~$x'$ as if the surface were uniform in the vicinity of that point. That approach corresponds to approximating~\eqref{eq:exact_rep_formula} by a similar equation, but with the exact Green's function $G$ replaced by an approximate ``proto"-Green's function $G_{\rm p}$ defined by the scattering of the source at $\ner'=(x',y')$ from a uniform surface $\alpha(x')$ and $\beta(x')$.  This approximation  yields 
\begin{equation}\label{eq:gp_tot}
 u^{\rm tot}_0(\ner) = u^\inc(\ner) +\int\displaylimits_{-\infty}^\infty\lf\{ G_{\rm p}(\ner|s,0^+)f^\inc_+(s)- G_{\rm p}(\ner|s,0^-)f^\inc_-(s)\rg\} \de s,\quad \ner\in \Omega_+\cup\Omega_-,
\end{equation}
which which turn out to be our zeroth order approximation in Sec.~\ref{eq:Somm_int}. We call $G_{\rm p}$ a proto-Green's function because it is a building-block for our solution, but it is not the Green's function one would get by putting a point source as the incident field in~\eqref{eq:gp_tot}.   We give an exact Green's function $G_{\rm p}$ for reflection and transmission off a uniform surface in Appendix~\ref{sec:Somm_int}.  But in the far field (fields far from the surface), as is derived rigorously in Appendix~\ref{sec:far_field}, this simplifies to a function $G_{\rm p}^{\rm ff}$ that  we present in a more elementary fashion here.

\begin{figure}[h!]
\centering	
\includegraphics[scale=0.9]{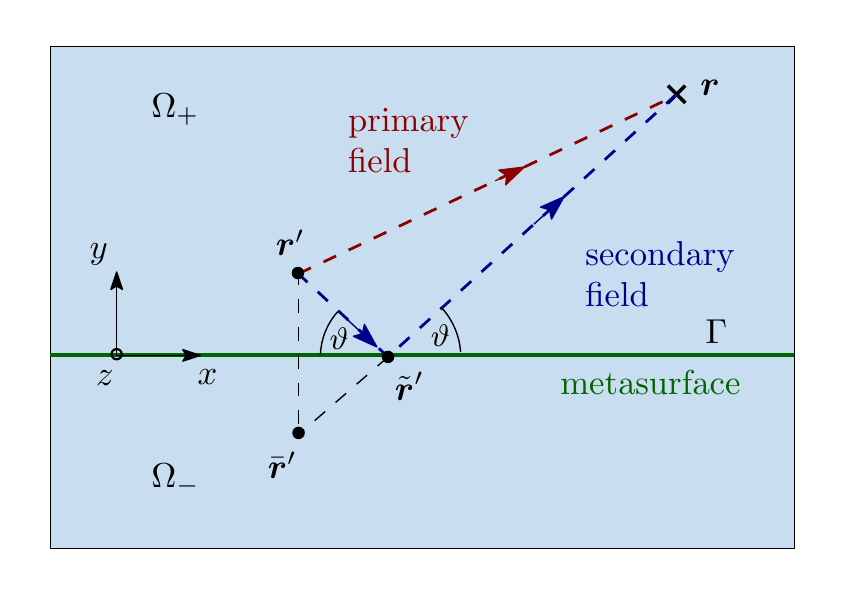}
\caption{Variables used in the derivation of the approximation~\eqref{eq:trans_coeff_approx} of the Green's function of the problem under consideration~\eqref{eq:huygens_Green}. The field produced by a point source at $\ner'$ impinges on the metasurface at $\tilde \ner'$. The total field is observed at the point $\ner$ far away from the metasurface that is denoted by $\Gamma$. The total field is decomposed in primary and secondary fields. The latter corresponds to the field  reflected off of the metasurface, which is approximated by means of the local reflection coefficient $R$ in~\eqref{eq:ref_coeff_approx} assuming specular reflection.}\label{fig:approx_GF}
\end{figure}

In order to construct $G_{\rm p}^{\rm ff}$, we consider the scattering configuration depicted in Fig.~\ref{fig:approx_GF}. 
With reference to that figure, the total wave field observed above the metasurface at a point  $\ner=(x,y)$, $y>0$ is given by the superposition $G^{\rm ff}_{\rm p} = G^\inc+\widetilde G^r$ of the (primary) incident field  $G^\inc(\ner|\ner') =\frac{i}{4}H_0^{(1)}(k|\ner-\ner'|)$ produced by a point source placed above the metasurface, at $\ner'=(x',y')$, $y'>0$, and the (secondary) field $\widetilde G^r$ resulting from the reflection at  $\tilde\ner'=(\tilde x',0)$ (on $\Gamma$)  of the ray stemming from $\ner'$. (The function $H_0^{(1)}$ is the Hankel function of first kind and order zero~\cite{abramowitz1964handbook}.) The magnitude and phase of the reflected field are characterized by the local \emph{reflection coefficient} $R(k\cos\theta,\tilde x')$ that depends on the reflection angle $\vartheta=\arctan(y/(x-\tilde x'))$ (which is measured with respect to tbe metasurface) of the reflected ray at~$\tilde \ner'$ (Fig.~\ref{fig:approx_GF}).  As is shown  in Appendix~\ref{sec:Somm_int}, the local reflection coefficient is given by 
\begin{equation}\label{eq:ref_coeff_approx}
 R( k_x,s) :=  1-\widehat\Gamma_\alpha(k_x,s)-\widehat\Gamma_\beta(k_x,s),
\end{equation} where 
\begin{equation}
\widehat\Gamma_\alpha(k_x,s) :=  \frac{k\alpha( s)}{k_y+k\alpha( s)},\quad \widehat\Gamma_\beta(k_x,s):=\frac{k\beta( s)}{k_y+k\beta(s)}\andtext k_y = \sqrt{k^2-k_x^2}.\label{eq:terms_RT}
\end{equation}
  In order to account for both the magnitude and the direction of the reflected field, we consider the image point source $\bar \ner' = (x',-y')\in\Omega_-$ which allows the total field field above the metasurface to be  approximated~as
\begin{subequations}\begin{equation}
 G^{\rm ff}_{\rm p}(\ner|\ner') = \underbrace{\frac{i}{4} H_0^{(1)}(k|\ner-\ner'|)}_{\rm incident\ field}+ \underbrace{\frac{i}{4}R(k\cos\vartheta,\tilde x')H_0^{(1)}(k|\ner-\bar\ner'|)}_{\rm reflected\ field}\qquad (\ner\in\Omega_+).\label{eq:ref_0}
\end{equation}

Similarly, below the metasurface the total field at a point~$\ner\in\Omega_-$ corresponds to the transmitted field which can be approximated as
\begin{equation}\label{eq:trans_0}
 G^{\rm ff}_{\rm p}(\ner|\ner')=  \frac{i}{4}T(k\cos\vartheta,\tilde x')H_0^{(1)}(k|\ner-\ner'|)\qquad (\ner\in\Omega_-),
\end{equation}\label{eq:KA_approx}\end{subequations}
in terms of the local \emph{transmission coefficient} which is shown in Appendix~\ref{sec:Somm_int} to be given by 
\begin{equation}\label{eq:trans_coeff_approx}
 T( k_x,s) := -\widehat\Gamma_\alpha(k_x,s)+\widehat\Gamma_\beta(k_x,s).
\end{equation}
in terms of $\widehat\Gamma_\alpha$ and $\widehat\Gamma_\beta$ defined in~\eqref{eq:terms_RT}. 

{For a point source placed below the metasurface at a point~$\ner'$ in $\Omega_-$, the approximate Green's function $G^{\rm ff}_{\rm p}$ can be derived from symmetry arguments.} A rigorous derivation of $G^{\rm ff}_{\rm p}$ based on  asymptotic analysis is presented in Appendix~\ref{sec:far_field}. 

{Note that $G^{\rm ff}_{\rm p}$ provides a valid approximation of the exact Green's function $G$ in the far-field zone, and can be thought of as the approximate field scattered from a single \emph{point} on the surface. In view of the Huygens' principle, to get the total field we must add together all of the surface points resulting in what we refer to as the \emph{ray-optics approximation}:}
\begin{equation}\label{eq:kirchhoff_approx}
 u^{\rm tot,ff}_{0}(\ner) = u^\inc(\ner) +\int\displaylimits_{-\infty}^\infty\lf\{ G^{\rm ff}_{\rm p}(\ner|s,0^+)f^\inc_+(s)- G^{\rm ff}_{\rm p}(\ner|s,0^-)f^\inc_-(s)\rg\} \de s,\quad \ner\in \Omega_+\cup\Omega_-,
\end{equation}
of the total field $u^{\rm tot}$ is achieved by replacing $G$ by $G^{\rm ff}_{\rm p}$ in the integral representation formula~\eqref{eq:exact_rep_formula}, where the functions $f^\inc_\pm$ are defined in~\eqref{eq:RHS} and where the limits $G^{\rm ff}_{\rm p}(\ner|s,0^\pm) = \lim_{\delta\to 0^+} G^{\rm ff}_{\rm p}(\ner|s,\pm\delta)$ are obtained by setting
\begin{equation}
R\lf(\frac{k(x-x')}{|\ner-\ner'|},x'\rg)\andtext T\lf(\frac{k(x-x')}{|\ner-\ner'|},x'\rg)\end{equation}
in formulae~\eqref{eq:ref_0} and~\eqref{eq:trans_0}, respectively. 

We will  show in Section~\ref{eq:Somm_int} that~\eqref{eq:kirchhoff_approx} is a simplified version of~\eqref{eq:gp_tot} in the limit of sources and fields far from the surface and, furthermore, that~\eqref{eq:kirchhoff_approx} and~\eqref{eq:gp_tot} are the zero-th order terms in a convergent series of corrections for slowly varying surfaces.

\begin{figure}[h!]
\centering	
\includegraphics[scale=0.45]{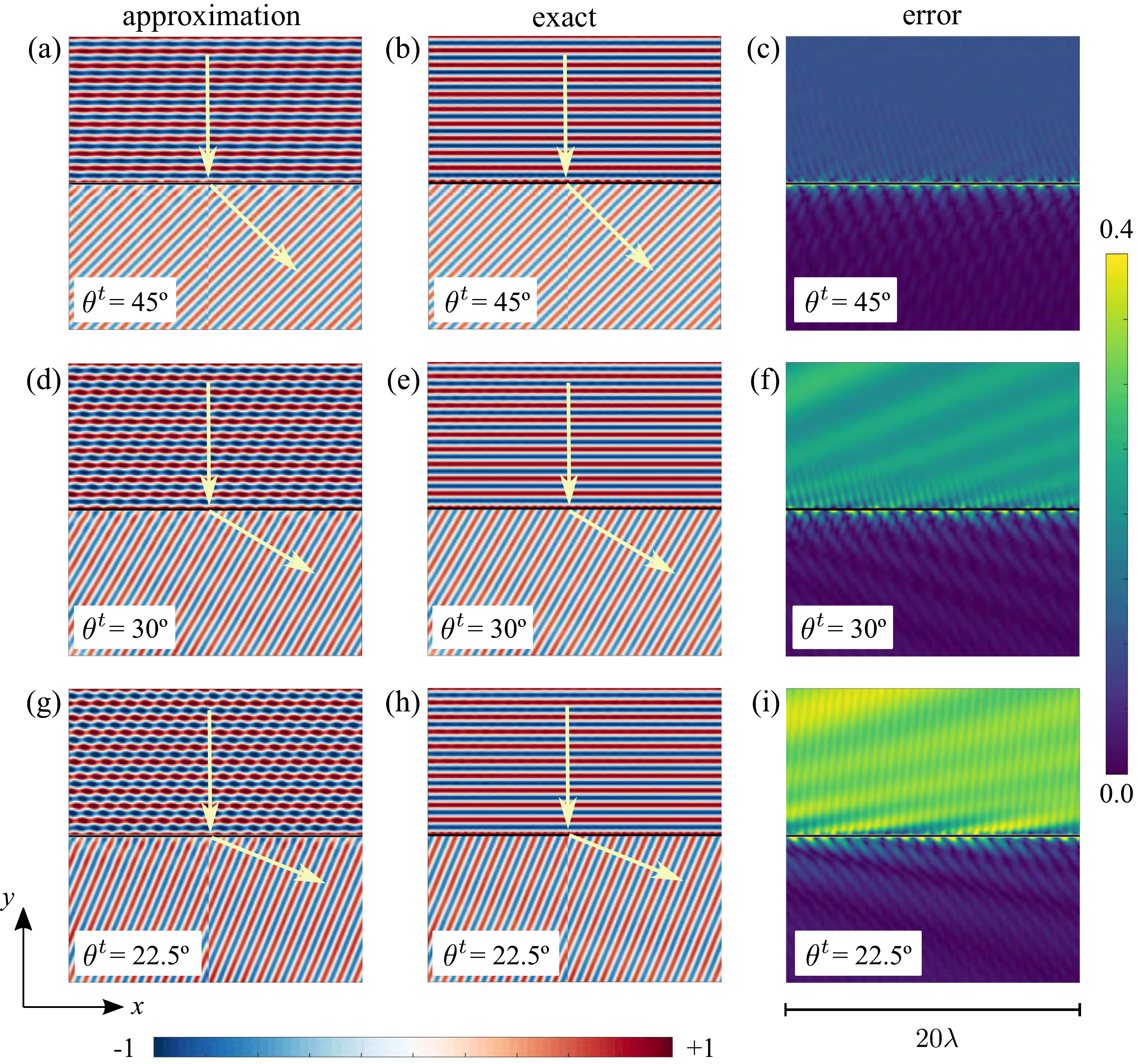}\\
\caption{First column: Ray-optics approximation~\eqref{eq:kirchhoff_approx} of the total field $u^{\rm tot}=u^\inc+u^{\rm scat}$ solution of the problem of scattering of a planewave that impinges at normal incidence on metasurface's that produce transmitted fields that are predominantly planewaves propagating in the directions $\theta^t$  (a)~$45^{\circ}$, (d)~$30^{\circ}$  and (g)~$22.5^{\circ}$ with respect to the metasurface. Second column: Full ``exact" total field solution of the corresponding for planewave transmitted fields in the directions $\theta^t$  (b)~$45^{\circ}$, (e)~$30^{\circ}$ and (h)~$22.2^{\circ}$. Third column: Absolute errors for planewave transmitted fields in the directions $\theta^t$   (c)~$45^{\circ}$, (f)~$30^{\circ}$ and (i)~$22.5^{\circ}$.}\label{fig:ex_1}
\end{figure}

In order to examine the accuracy of the ray-optics approximation of the total-field~\eqref{eq:kirchhoff_approx}, we consider a series of numerical examples.  Fig.~\ref{fig:ex_1}  presents a comparison of the approximate and ``exact" total field solution of the problem of scattering of a planewave that impinges at normal incidence upon three different mesaturfaces. The metasurface parameters $\alpha$ and $\beta$ were selected so that the transmitted fields satisfy the so-called \emph{generalized laws of reflection and transmission}~\cite{Yu:2011eya,Yu:2014hq,Epstein:2014bx}. {As is shown in Appendix~\ref{app:GLRT}},  for a given incident planewave $u^\inc$ in the direction $\hat{\bold k}^\inc=(\cos\theta^\inc,\sin\theta^\inc)$, $-\pi/2\leq\theta^\inc \leq 0$, metasurface parameters of the form
\begin{subequations}\begin{equation}\alpha(x) = c_0\lf(1\pm\e^{ik xd}\rg)\andtext\beta(x) = c_0\lf(1\mp\e^{ikxd}\rg),\end{equation}
with 
\begin{equation}c_0 = -\sin\theta^\inc \andtext d = \cos\theta^ t-\cos\theta^\inc,\end{equation}\label{eq:GLRT}\end{subequations}
produce (to leading-order asymptotics) a transmitted field corresponding to a single planewave in the direction $\hat{\bold k}^t=(\cos\theta^t,\sin\theta^t)$.

Although the ray-optics approximation~\eqref{eq:kirchhoff_approx}  seems to capture qualitatively the main features of the scattered field, quantitatively it exhibits large near-field errors---and also large far-field errors in some cases~(Fig.~\ref{fig:ex_1}(i), for example)---that are just one order of magnitude smaller than the scattered field itself.  In what follows of this paper we present a methodology to produce both near- and far-field corrections to the ray-optics approximation~\eqref{eq:kirchhoff_approx}, which turns out to be just the zeroth-order terms of a series  approximation of the exact scattered field in the far-field.

\section{Locally uniform approximation and corrections to ray-optics}\label{eq:Somm_int}

This section presents an SIE formulation of the problem of scattering~\eqref{eq:huygens_scattering} from which corrections to the ray-optics approximation~\eqref{eq:kirchhoff_approx} can be easily obtained in the form of a Born (or Neumann) series~\cite{born1999principles}. We here follow a standard indirect integral equation formulation procedure~\cite{colton2013integral,nedelec2001acoustic} in which the field is represented by means of a Green's function $G_{\rm p}$ that satisfies the Helmholtz equation in both upper and lower homogeneous media, but does not satisfy the correct sheet transition conditions at  the metasurface, and we solve for effective source terms that restore the desired transition conditions. Note that, unfortunately, $G_{\rm p}^{\rm ff}$ cannot be used for this purpose because it does not satisfy the Helmholtz equation, since both reflection and transmission coefficients depend on the location of the source and observation points.


Just as in Section~\ref{sec:k_approx}, we begin by constructing a \emph{proto-Green's function} $G_{\rm p}$, which is given in terms of Fourier-like integrals that we deem as \emph{Sommerfeld integrals} (due to the similarities they share with  layered-media Sommerfeld integrals~\cite{chew1995waves}). This Green's function possesses two important features. On one hand $G_{\rm p}$---as $G$ itself---satisfies the inhomogeneous Helmholtz equation~\eqref{eq:Helm_PS} with a point source excitation and, on the other hand, its far field equals the approximation~$G^{\rm ff}_{\rm p}$ of the exact Green's function $G$, used in the ray-optics approximation~\eqref{eq:kirchhoff_approx}. The former allows us to properly derive a second-kind SIE, while the latter guarantees that the zeroth-order approximation obtained by truncation of the Born series solution of the second-kind integral equation does indeed correspond to the ray-optics approximation~\eqref{eq:kirchhoff_approx} in the far-field zone. As in Section~\ref{sec:k_approx}, we use the term ``proto" because $G_{\rm p}$ is just a building block  and  is not equal to the Green's function of our final zeroth-order approximation.

The key feature of the proto-Green's function is that, instead of satisfying the non-local transition conditions~\eqref{eq:BC_alpha}--\eqref{eq:BC_beta}, it satisfies the following \emph{local transition conditions}
{\begin{equation}
\ljmp G_{{\rm p},y}(\ner|\ner')\rjmp=-ik\alpha(x')\lavg  G_{\rm p}(\ner|\ner')\ravg\ \mbox{ and }\ \lavg G_{{\rm p},y}(\ner|\ner')\ravg= -ik\beta(x')\ljmp  G_{\rm p}(\ner|\ner')\rjmp,\ \ner\in\Gamma,\label{eq:approx_Green}
\end{equation}}
with metasurface parameters $\alpha$ and $\beta$~depending on the source point $\ner'=(x',y')$. This local transition conditions formalize the slowly varying assumption used in~Section~\ref{sec:k_approx} where the ray-optics approximation was derived. Since the metasurface parameters $\alpha$ and $\beta$ that appear in local transition conditions~\eqref{eq:approx_Green} do not depend on the observation point $\ner=(x,y)$, they can be treated as constants and, thus, an analytical expression for $ G_{\rm p}$  in terms of Sommerfeld integrals can be easily obtained. The idea behind this calculation is to decompose the point-source incident field $G^\inc$ as a superposition of both propagative and evanescent planewaves. Since specular reflection takes place for each plane wave impinging on the metasurface, the resulting scattered field can be written down as a superposition of reflected and transmitted plane waves weighted by the reflection and transmission coefficients provided in~\eqref{eq:ref_coeff_approx}  and~\eqref{eq:trans_coeff_approx}, respectively.  The details of this derivation are presented in Appendix~\ref{sec:Somm_int}. Furthermore,  it is shown in Appendix~\ref{sec:far_field} by means of a detailed asymptotic analysis that, to leading asymptotics, $G_{\rm p}$ equals $G^{\rm ff}_{\rm p}$ as $|\ner|\to\infty$.

With an analytical expression for $G_{\rm p}$ in hand (i.e., formulae~\eqref{eq:G_up} and~\eqref{eq:G_dw}) we proceed to derive an SIE for the solution of the scattering problem~\eqref{eq:huygens_scattering} from which corrections to the ray-optics approximation~\eqref{eq:kirchhoff_approx} can be computed. Following the exact integral representation~\eqref{eq:exact_rep_formula} of the scattered field $u^{\rm scat}$ we  introduce an indirect integral formulation for the scattering problem~\eqref{eq:huygens_scattering} by setting
\begin{equation}\label{eq:int_rep}
 u^{\rm scat}(\ner)= \int\displaylimits_{-\infty}^\infty\lf\{G_{\rm p}(\ner|\sigma,0^+)\varphi(\sigma)-G_{\rm p}(\ner|\sigma,0^-)\psi(\sigma)\rg\}\de \sigma,\quad \ner\in \Omega_+\cup\Omega_-,
\end{equation}
where $\varphi$ and~$\psi$ are (so far) unknown surface density functions.  Note that if $G_{\rm p}$ were the exact Green's function $G$, then $u^{\rm scat}$ in~\eqref{eq:int_rep} would be the exact solution of~\eqref{eq:huygens_scattering} provided $\varphi=f^\inc_+$ and $\psi=f^\inc_-$, where $f^\inc_+$ and $f^\inc_-$ are defined in~\eqref{eq:RHS}.  Note further that in virtue of the relationship between $G_{\rm p}$ and $G^{\rm ff}_0$ established in Appendix~\ref{sec:far_field}, the substitutions $\varphi=f^\inc_+$ and $\psi=f^\inc_-$ in~\eqref{eq:int_rep} would produce an approximation of $u^{\rm scat}$ that, in the far-field zone, exhibits the same accuracy of the ray-optics approximation~\eqref{eq:kirchhoff_approx}.  

Continuing with the derivation of the SIE, we observe that in order for $u^{\rm scat}$~\eqref{eq:int_rep} to be an exact solution of~\eqref{eq:huygens_scattering} it has to satisfy both the Helmholtz equation~\eqref{eq:HEqn} and the transition conditions~\eqref{eq:bc_1}-\eqref{eq:bc_2}. The problem here is that,  although $u^{\rm scat}$~\eqref{eq:int_rep} does satisfy the Helmholtz equation~\eqref{eq:HEqn} in $\Omega_+$ and $\Omega_-$ for any admissible densities $\varphi$ and $\psi$ (since be construction $G_{\rm p}(\ner|\ner')$, $\ner'\in\Gamma$, satisfies it), it does not necessarily fulfill the correct transition conditions~\eqref{eq:BC_alpha}--\eqref{eq:BC_beta} unless $(\varphi,\psi)$ is solution of a certain SIE.  Indeed, it is shown in Appendix~\ref{sec:integral_eq_corrections} that, imposing the transition conditions~\eqref{eq:BC_alpha}--\eqref{eq:BC_beta} on $u^{\rm scat}$ in~\eqref{eq:int_rep}, an SIE for the unknown density functions $(\varphi,\psi)$ is obtained. The resulting equations correspond to two decoupled second-kind SIEs:
\begin{eqnarray}
\mu_j -\operatorname T_j[\mu_j] = g^\inc_j, \qquad j=1,2, 
\label{eq:system_abstract}\end{eqnarray}
for the unknown auxiliary densities  $\mu_j$, $j=1,2$, which are directly  related to the integral densities in~\eqref{eq:int_rep} by 
\begin{equation}\label{eq:rel_mu_phi_psi}
\varphi = \frac{\mu_1+\mu_2}{2} \andtext\psi=\frac{\mu_2-\mu_1}{2}.\end{equation}
 The precise definition of the integral operators $\operatorname T_j$, $j=1,2,$ is given in Appendix~\ref{sec:integral_eq_corrections} and  the functions $g^\inc_j$, $j=1,2$, on right-hand-side of~\eqref{eq:system_abstract}, are
\begin{equation}
g^\inc_1(s) = 2ik\alpha(s) u^\inc(s,0)\andtext g^\inc_2(s) = 2u^\inc_y(s,0).\label{eq:aux_rhs}\end{equation}

Clearly, the densities  $\mu_j$, $j=1,2,$ can be determined by solving the SIEs~\eqref{eq:system_abstract} and from them, the desired densities $\varphi$ and $\psi$ that make $u^{\rm scat}$ in~\eqref{eq:int_rep} the exact solution of~\eqref{eq:huygens_scattering} can be readily obtained.

We now recall that we are here interested in slowly varying interface parameters $\alpha$ and $\beta$ of the form $\alpha(x)=a(\varepsilon x)$ and $\beta(x)=b(\varepsilon x)$ where $\varepsilon>0$ is a small parameter. In view of  definitions~\eqref{eq:int_ker_IE} and~\eqref{eq:IOps}, we  observe  that both integral operators, $\operatorname T_1$ and $\operatorname T_2$, vanish as  $\varepsilon\to0$ and, therefore, in the limit when  $\varepsilon=0$ the exact SIE solutions are simply $\mu_j=g^\inc_j$, $j=1,2$. For small but nonzero values of $\varepsilon$, in turn, convergent Neumann-series solutions
\begin{equation}
\mu_j = \sum_{n=0}^\infty \operatorname  T_j^ng^\inc_j,\qquad j=1,2,\label{eq:neu_ser}
\end{equation}
 of the SIEs~\eqref{eq:system_abstract} can be obtained because the integral operators satisfy  $\|\operatorname T_j\|<1$ in a certain operator norm for sufficiently small $\varepsilon$.

 The $N$th-order approximations of the density functions $\varphi$ and $\psi$ can thus be defined as
\begin{equation}\varphi_N = \frac{\mu^{(N)}_1+\mu^{(N)}_2}{2}\andtext\psi_N = \frac{\mu_2^{(N)}-\mu_1^{(N)}}{2},\label{eq:relation}\end{equation}
where $\mu_j^{(N)}$, $j=1,2$, are  the truncated Neumann series
\begin{equation}
\mu_j^{(N)}(s) :=
\sum_{n=0}^N \operatorname T_j^ng^\inc_j(s),\quad j=1,2,\quad N\geq 0. \label{eq:trunc_Neu_series}
\end{equation}
From~\eqref{eq:relation} we then  define the $N$th-order locally uniform approximation of the total near and far fields:
\begin{subequations}\begin{eqnarray}\label{eq:int_rep_ord}
u^{\rm tot}_N (\ner)&:=&u^\inc (\ner)+ \int\displaylimits_{-\infty}^\infty\lf\{G_{\rm p}(\ner|\sigma,0^+)\varphi_N(\sigma)- G_{\rm p}(\ner|\sigma,0^-)\psi_N(\sigma)\rg\}\de \sigma\quad\mbox{and}\label{eq:int_rep_ord}\\
u^{{\rm tot,ff}}_N (\ner)&:=&u^\inc (\ner)+ \int\displaylimits_{-\infty}^\infty\lf\{G_{\rm p}^{\rm ff}(\ner|\sigma,0^+)\varphi_N(\sigma)- G_{\rm p}^{\rm ff}(\ner|\sigma,0^-)\psi_N(\sigma)\rg\}\de \sigma,\label{eq:int_rep_ord_ff}
\end{eqnarray}\label{eq:corrections}\end{subequations}
 for $ \ner\in \R^{2}\setminus\Gamma$, respectively. {Note the zeroth-order term (or any higher order term) in~\eqref{eq:corrections} does \emph{not} correspond to setting $\varepsilon=0$ (a uniform surface).}
 
Finally, it follows from the definitions above that the ray-optics approximation~\eqref{eq:kirchhoff_approx} is simply the zeroth-order approximate far-field $u_{\rm 0}^{\rm tot,ff}$, i.e., the $N=0$ instance of the formula~\eqref{eq:int_rep_ord_ff}. In order to see this it suffices to note that $\varphi_0 = f^\inc_+$ and $\psi_0=f^\inc_-,$
which followed directly the definition of $f^\inc_\pm$ in~\eqref{eq:RHS} and  the fact that 
\begin{equation}
\mu_1^{(0)}(s) :=g^\inc_1(s)= 2ik \alpha u^\inc(s,0)\andtext \mu^{(0)}_2(s) :=g^\inc_2(s)= 2u^\inc_y(s,0).
\end{equation}

In summary, the $N$th-order approximation of the total near and far fields resulting from the scattering of an incident field $u^\inc$ off of a metasurface $\Gamma$, can obtained as follows:
\begin{enumerate}
\item Evaluate the input data $g_j^\inc$, $j=1,2,$ defined in~\eqref{eq:aux_rhs}, using the prescribed incident field~$u^\inc$.
\item Compute the $N$th-order approximate densities $\mu_j^{(N)}$, $j=1,2,$ defined in~\eqref{eq:trunc_Neu_series}, by repeated application of the integral operators $\operatorname T_j$ to $g_j^\inc$. 
\item Evaluate the approximate densities $\varphi_N$ and $\psi_N$, defined in~\eqref{eq:relation}, by taking suitable linear combinations of $\mu_j^{(N)}$, $j=1,2$, obtained in step 2.
\item To produce the approximate near (resp. far) field, substitute the densities $\varphi_N$ and $\psi_N$ obtained in step~3 in the integral formula~\eqref{eq:int_rep_ord} (resp.~\eqref{eq:int_rep_ord_ff}). 
\end{enumerate}

\begin{figure}[h!]
\centering	
\includegraphics[scale=0.44]{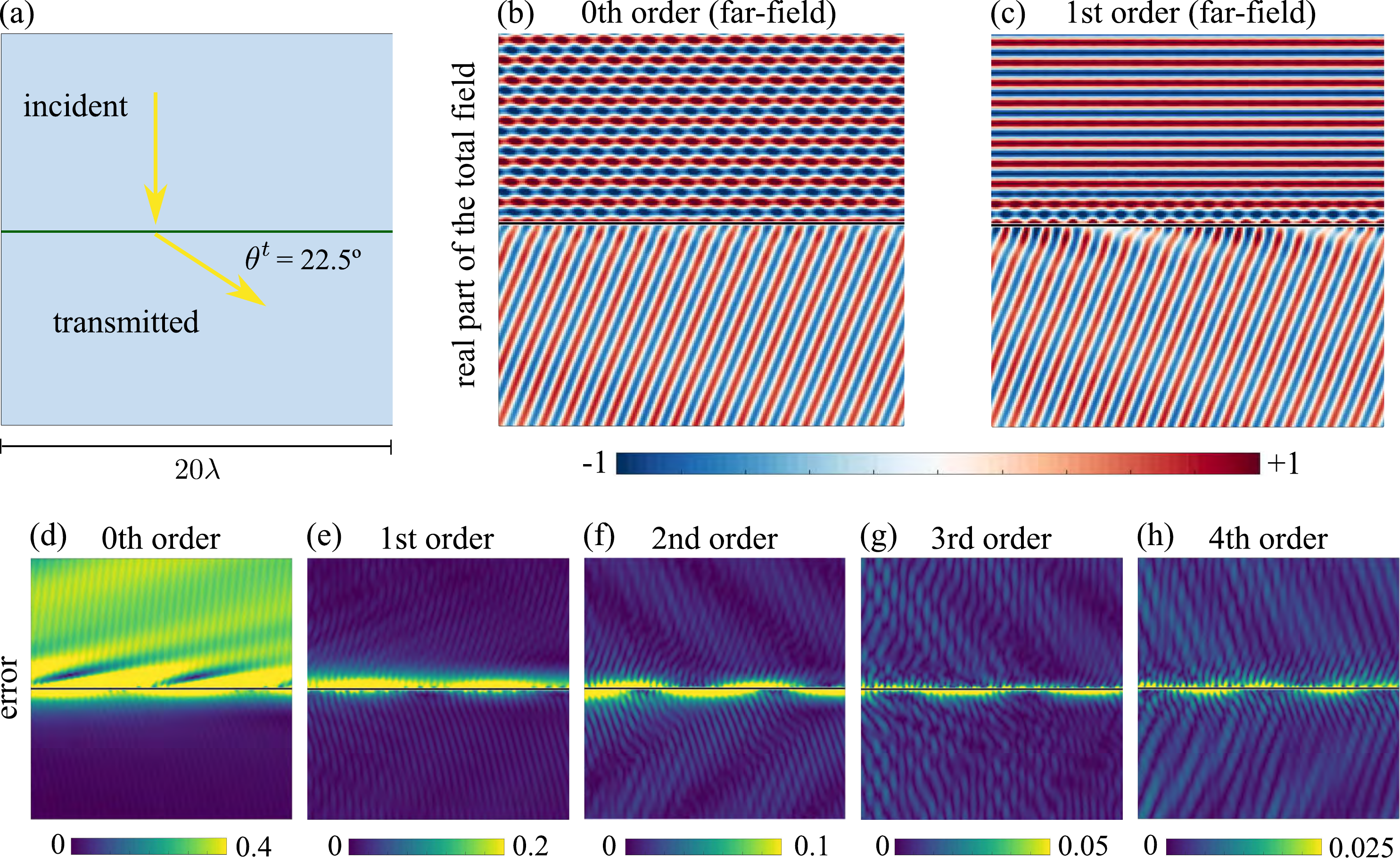}
\caption{High-order corrections to the ray optics approximation~\eqref{eq:kirchhoff_approx}. (a) Geometrical configuration of the problem under consideration which corresponds to an unit amplitude planewave impinging at normal incidence on a metasurface that renders a transmitted planewave with wavevector forming an angle of $22.5^{\circ}$ with respect to the metasurface. (b) and (c): Real part of the total field 0th and 1st order approximations. (d), (e), (f), (g) and (h): Absolute errors $|u-u^{\rm tot,ff}_N|$, for $N=0,1,2,3$ and~$4$, respectively, in the zeroth (no correction), first, second, third and fourth order corrections to the ray optics approximation $u_0^{\rm tot,ff}$. The color scales were adjusted according to the maximum error displayed in each one of the figures.}\label{fig:ex_2}
\end{figure}

In order to illustrate the accuracy yielded by the higher-order corrections to the ray-optics (zeroth-order) approximation, we  present  Fig.~\ref{fig:ex_2} which concerns the scattering configuration considered above in Figs.~\ref{fig:ex_1}(g)--(i), which corresponds to the scattering of a planewave that impinges at normal incidence on a metasurface that renders a transmitted planewave with wavevector forming an angle of $22.5^\circ$ with respect to the metasurface (Fig.~\ref{fig:ex_2}(a)).  Figs.~\ref{fig:ex_2}(b) and~\ref{fig:ex_2}(c) display the real part of the total fields (incident + reflected,  and transmitted) produced by the zeroth and first order  approximations of the far field, corresponding to formula~\eqref{eq:int_rep_ord_ff} with $N=0,1$. 
 In order to better visualize the convergence of the locally uniform approximations~\eqref{eq:int_rep_ord_ff} as $N$ increases, we present Figs.~\ref{fig:ex_2}(d)--(h) that display the absolute value of the zeroth, first, second, third and fourth order far-field errors. The reference ``exact" far-field was computed by direct solution of the SIE~system~\eqref{eq:system_abstract}. These results  indicate that the far-field error is roughly reduced by a factor of~0.5 as the order increases. This is explained by the  fact that the spectral radii of the discrete versions of the integral operators  $\operatorname T_j$, $j=1,2,$ are approximately $0.5$. More details on the convergence of the Neumann series approximation are given in Appendix~\ref{sec:conv_neu_ser}.

\begin{figure}[h!]
\centering	
\includegraphics[scale=0.45]{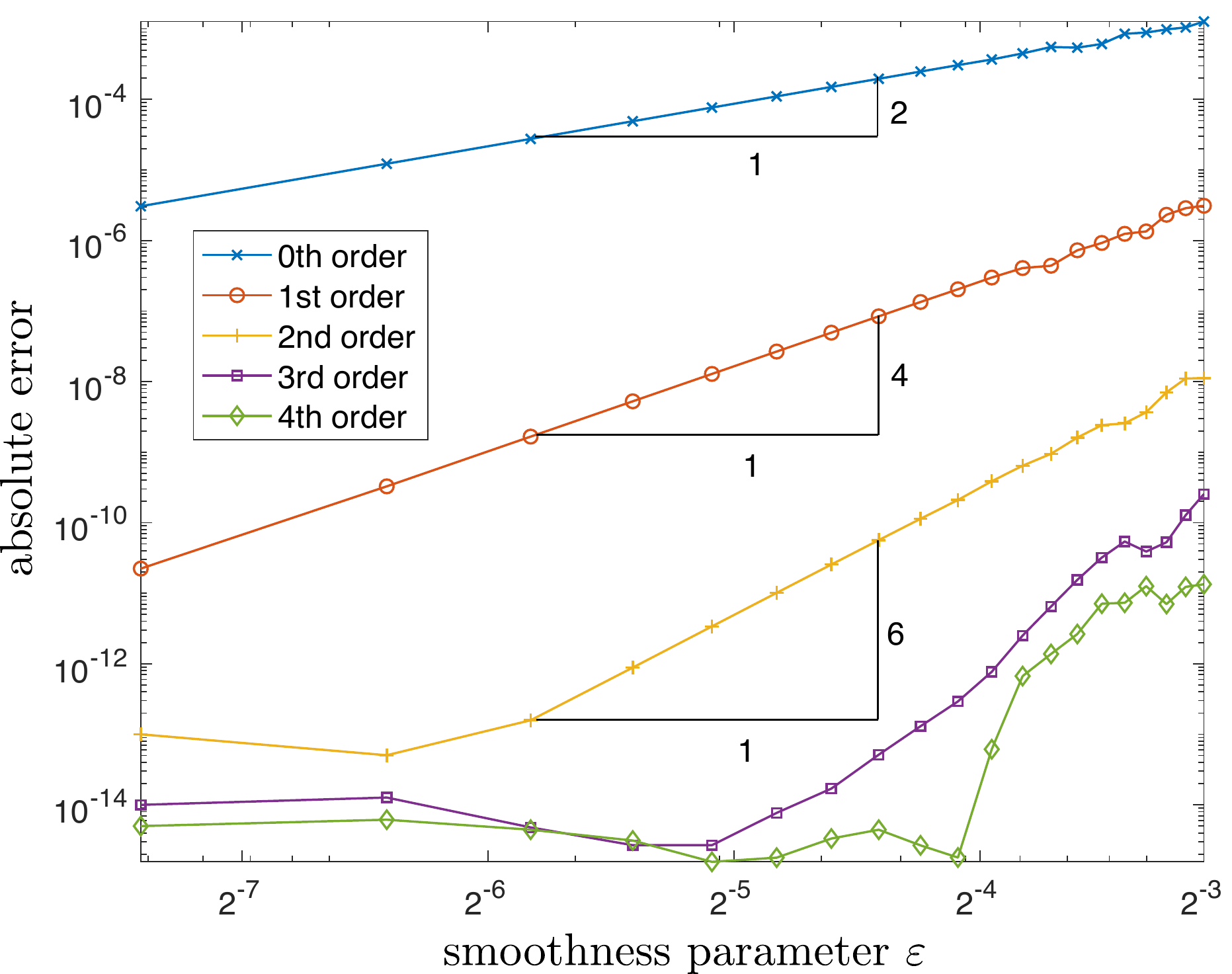}
\caption{Rate of convergence of the zeroth-, first-, second-, third- and fourth-order near-field approximations~\eqref{eq:int_rep_ord} in terms of the smoothness of the metasurface parameters. The plot is in log-log scale. The problem under consideration is the scattering of a unit amplitude plane wave at normal incidence that impinges on a metasurface with metasurface parameters  $\alpha(x) =a(\varepsilon x)=c_0\{1-\e^{i\varepsilon x}\}$ and $\beta(x) =b(\varepsilon x) =c_0\{1-\e^{-i\varepsilon x}\}$. The parameter $\varepsilon>0$ controls the smoothness of  metasurface parameters. The color curves display the maximum of the absolute error $|u^{\rm tot}(\ner)-u^{\rm tot}_{N}(\ner)|$ in the near-field approximations~\eqref{eq:int_rep_ord} evaluated at the spatial points $\ner\in\{-1,0,1\}\times\{-10,10\}$.} \label{fig:ex_convergence}
\end{figure}

Our next example concerns the dependence of the rate of convergence of the $N$th-order approximations~\eqref{eq:int_rep_ord} and~\eqref{eq:int_rep_ord_ff} on the smoothness of the metasurface parameters. As it turns out,  for constant metasurface parameters the zeroth-order near and far field approximations are exact. In this example we thus attempt to quantify how errors depart from zero as the metasurface parameters become non-constant. In order to so we consider slowly-varying metasurface parameters $\alpha(x)=a(\varepsilon x)$ and $\beta=b(\varepsilon x)$---which depend on a small parameter $\varepsilon>0$---that tend to constants $\alpha_0=a(0)$ and $\beta_0=b(0)$ as $\varepsilon\to 0$. Fig.~\ref{fig:ex_convergence} displays the near-field errors for vanishing values of the smoothness parameter $\varepsilon>0$. Clearly, the zeroth-, first and third-order approximations exhibit errors of order $O(\varepsilon^2)$, $O(\varepsilon^4)$ and $O(\varepsilon^6)$, respectively, as $\varepsilon\to 0$, i.e., as the metasurface parameters tend to constants.

\begin{figure}[h!]
\centering	
\includegraphics[scale=0.41]{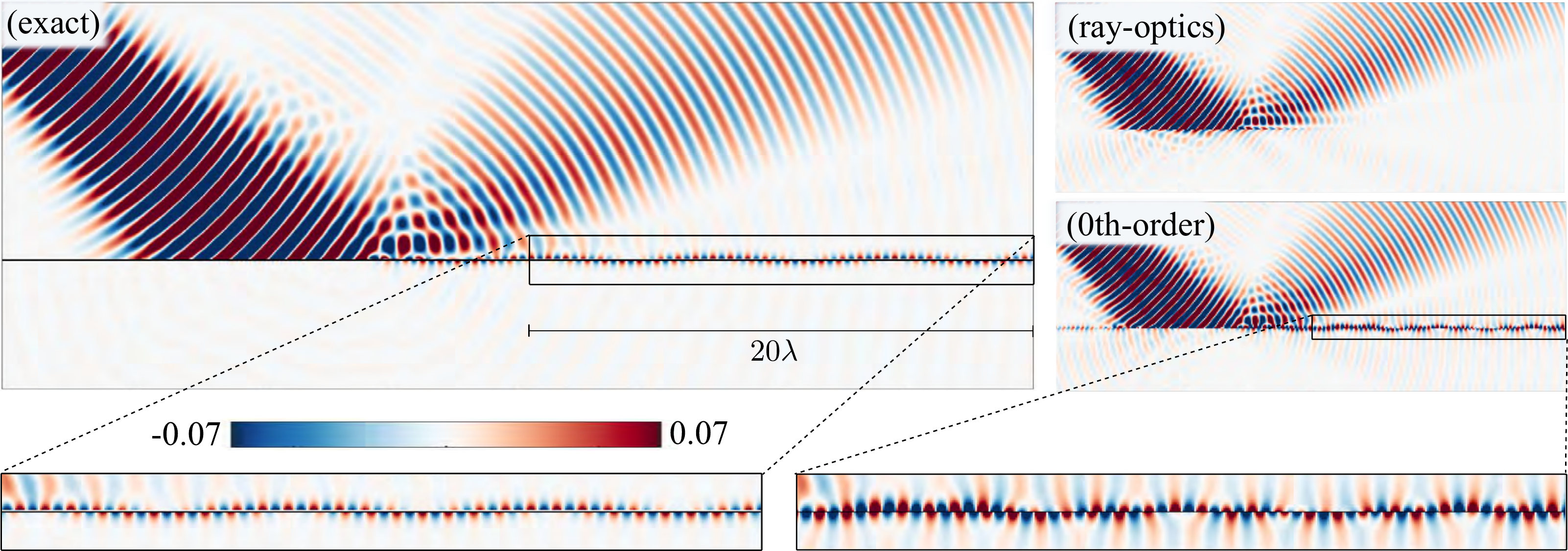}\\
\caption{Total field solution of the problem of scattering of a Gaussian beam by a metasurface with interface parameters $\alpha$ and $\beta$ engineered to allow for surface wave modes~\eqref{eq:surf_wave_mode} to propagate along the metasurface.}\label{fig:ex_beam}
\end{figure}

Interestingly, the detailed asymptotic calculations presented in Appendix~\ref{sec:far_field} also reveal that surface-wave modes appear in the asymptotic expansion of $ G_{\rm p}$ for certain constant values of the metasurface parameters (note that for constant $\alpha$ and $\beta$, it holds that $G=G_{\rm p}$). Such surface-wave modes are also present in the field scattered by metasurfaces with non-constant metasurface parameters. To demonstrate this fact, we present Fig.~\ref{fig:ex_beam} which displays the total field solution of the problem of scattering of a Gaussian beam by a metasurface for which a surface-wave mode propagates from left to right along. Three difference solution are displayed in that figure: the exact solution, the ray optics approximation~\eqref{eq:kirchhoff_approx}, and the zeroth-order locally uniform approximation~\eqref{eq:int_rep_ord} with $N=0$. Since $G_{\rm p}^{\rm ff}$ is a far-field approximation (which is valid at a certain distance from the metasurface), $u_0^{\rm tot,ff}$ does not capture at all  the aforementioned surface-wave modes.

\section{Concluding remarks}\label{sec:conclusions}
We developed an SIE approach, based on a locally uniform approximation of a metasurface, to establish the accuracy and compute higher-order corrections to a ray-optics approximation commonly used in inverse metasurace design, where metasurfaces are modeled by means of slowly varying surface parameters.

This work opens many research directions that could be pursued in the future. The most important (and straightforward, in principle) is perhaps the extension of the proposed approach to three-spatial dimensions. As a practical matter, however, there are many subtle implementation aspects of this extension, such as the derivation of suitable three-dimensional SIE formulations and the efficient evaluation of the resulting two-dimensional surface integrals, that need to be addressed.

Another future research direction is the extension of the proposed approach to more general classes of metasurfaces that cannot be modeled by means of sheet transition conditions and in particular, to approximate metasurfaces as locally periodic rather than locally uniform. This extension, however, poses new theoretical challenges. Such an extension requires the knowledge of a certain proto-Green's function associated with a periodic transmission problem, which does not admit an expression in terms of Sommerfeld integrals and must be  computed numerically. Despite these theoretical challenges, there is both numerical and experimental evidence that a locally periodic approximation is sufficiently accurate for practical metasurface design~\cite{Pfeiffer:2014bv,achouri2015general,Aieta:2012gm,yu2013flat,Yu:2011eya,Yu:2014hq,raphael_paper}.

Finally, we mention that there remains considerable room for further asymptotic analysis of the integral operators~$\operatorname T_j$, $j=1,2,$ and their convergence as $\varepsilon\to 0$, to rigorously establish the convergence rate of the Neumann-series solution~\eqref{eq:neu_ser} (i.e. the corrections to ray optics). Although it is clear that $\Vert \operatorname T_j \Vert \to 0$ as $\varepsilon \to 0$, an intricate analysis is required to obtain convergence rates, and to clearly specify for which function spaces convergence is obtained, especially for unbounded surfaces and incident fields where limiting processes are tricky to apply to the surface integrals~\eqref{eq:IOps}. 



%
%
%

\section*{Funding}
This work was supported in part by the Army Research Office and under Cooperative Agreement Number W911NF-18-2-0048.

\section*{Appendices}
\appendix

\section{Exact integral representation}\label{app:symmetry}
This appendix is devoted to the derivation of the integral representation formula~\eqref{eq:exact_rep_formula}. 
In order to achieve that, we show first the symmetry of the exact Green's function~\eqref{eq:huygens_Green}.  
Consider then the functions $w(\ner)=G(\ner|\ner_1)$ and $v(\ner)=G(\ner|\ner_2)$ for the $\ner_1,\ner_2\in\Omega_+$. By Green's formula and the radiation condition we have 
\begin{equation}
v(\ner_1)-w(\ner_2) = \int_{\Omega_+} \lf\{w\nabla^2 v - v\nabla^2 w\rg\}\de\ner =-\int\displaylimits_{-\infty}^\infty\lf\{w^+v^+_y-v^+w^+_y\rg\}\de s,\label{eq:symmetry_1}
\end{equation}
and, similarly 
\begin{equation}
0=\int_{\Omega_-} \lf\{w\nabla^2 v - v\nabla^2 w\rg\}\de\ner =\int\displaylimits_{-\infty}^\infty\lf\{w^-v^-_y-v^-w^-_y\rg\}\de s.\label{eq:symmetry_2}
\end{equation}
From the transition conditions~\eqref{eq:transition_conditions}, on the other hand, it follows that  
\begin{equation}
w^+v^+_y-v^+ w^+_y =w^-v^-_y-v^-w^-_y.\label{eq:imp_identity}
\end{equation} where $v^\pm(s) =v(s,0^\pm)=\lim_{\delta\to 0^\pm}v(s,\delta)$ and $ v_y^\pm(s) = v_y(s,0^\pm)=\lim_{\delta\to 0^\pm}v_y(s,\delta)$ and similarly for $w$. Combining~\eqref{eq:symmetry_1},~\eqref{eq:symmetry_2} and~\eqref{eq:imp_identity} it is  obtained that
$
v(\ner_1)-w(\ner_2)=G(\ner_1|\ner_2)-G(\ner_2|\ner_1) =0,
$
and thus $G(\ner|\ner')=G(\ner'|\ner)$ for all  $\ner,\ner'\in\Omega_+$. The identity $G(\ner|\ner')=G(\ner'|\ner)$ for all  $\ner,\ner'\in\Omega_-$ can be derived in a completely analogous way. Consider now the  functions $w(\ner)=G(\ner|\ner_1)$ and $v(\ner)=G(\ner|\ner_2)$ but with $\ner_1\in\Omega_+$ and $\ner_2\in\Omega_-$.  Integration by parts yields the identity
\begin{equation}
v(\ner_1)=  -\int\displaylimits_{-\infty}^\infty\lf\{w^+ v^+_y-v^+w^+_y\rg\}\de s = w(\ner_2)\label{eq:symmetry_3}
\end{equation}
in this case, which clearly implies that $G(\ner_1|\ner_2) = G(\ner_2|\ner_1)$. The same result can be easily obtained in the case $\ner_1\in\Omega_-$ and $\ner_2\in\Omega_+$. 

Finally, the identities
{\begin{equation}\begin{split}
\ljmp G_{y'}(\ner|\ner')\rjmp = -k\alpha(x')\{\!\{G(\ner|\ner')\}\!\}\andtext \lavg G_{y'}(\ner|\ner')\ravg = -k\beta(x')\ljmp G(\ner|\ner')\rjmp
\end{split}\label{eq:BC_GF_1}\end{equation}}
involving the normal derivatives of the Green's function on the metasurface $\Gamma$ for $\ner\in\Omega_+\cup\Omega_-$, follow straightforwardly from the reciprocity condition $G(\ner|\ner')=G(\ner'|\ner)$ established above.

Having established the symmetry of the Green's function,~i.e., $G(\ner|\ner') = G(\ner'|\ner)$ for all $\ner,\ner'\in \Omega_-\cup\Omega_+,$ we can now use it to prove the identity~\eqref{eq:exact_rep_formula}. Indeed, it follows from~\eqref{eq:huygens_scattering}, the symmetry of $G$, and Green's theorem  that
\begin{equation}\label{eq:above}
\int\displaylimits_{-\infty}^\infty\lf\{G_{y'}(\ner|s,0^+)u^{\rm scat,+}(s)-G(\ner|s,0^+)u^{\rm scat,+}_y(s)\rg\} \de s =\lf\{\begin{array}{cl}u^{\rm scat}(\ner),&\ner\in \Omega_+,\medskip\\
0,&\ner\in \Omega_-, \end{array}\rg.
\end{equation}
and 
\begin{equation}\label{eq:below}
-\int\displaylimits_{-\infty}^\infty\lf\{G_{y'}(\ner|s,0^-)u^{\rm scat,-}(s)-G(\ner|s,0^-)u^{\rm scat,-}_y(s)\rg\} \de s =\lf\{\begin{array}{cl}0,&\ner\in \Omega_+,\medskip\\
u^{\rm scat}(\ner),&\ner\in \Omega_-. \end{array}\rg.
\end{equation}
Writing the the normal derivatives of $G$ and $u^{\rm scat}$ in terms of their limit values from above and below~$\Gamma$, it follows that
\begin{equation}\begin{split}
G_{y'}(\ner|s,0^-) =&~A(s)G(\ner|s,0^-) +B(s)G(\ner|s,0^+),\\
 G_{y'}(\ner|s,0^+)=&~-B(s)G(\ner|s,0^-) -A(s)G(\ner|s,0^+),  \\
u^{\rm scat,-}_y(s) =&~A(s)u^{\rm scat,-}(s) +B(s)u^{\rm scat,+}(s)-f^\inc_-(s)\andtext \\
u^{\rm scat,+}_y(s) =&~-B(s)u^{\rm scat,-}(s) -A(s)u^{\rm scat,+}(s)-f^\inc_+(s),
\end{split}\end{equation}
where \begin{equation}
A(s) =ik \lf\{\frac{\alpha(s)+\beta(s)}{2}\rg\}\andtext B =ik \lf\{\frac{\alpha(s)-\beta(s)}{2}\rg\}
\end{equation} 
and $f^\inc_{\pm}$ defined in~\eqref{eq:RHS}. Appropriately combining these expressions we arrive at
\begin{align}
G_{y'}(\ner|s,0^-)u^{\rm scat,-}(s)-G(\ner|s,0^-)u_y^{\rm scat,-}(s)=&~G(\ner|s,0^-)f^\inc_-(s)+\\
&B(s)\lf(G(\ner|s,0^+) u^{\rm scat,-}(s )-G(\ner|s,0^-)u^{\rm scat,+}(s)\rg) \nonumber\\
G_{y'}(\ner|s,0^+)u^{\rm scat,+}(s)-G(\ner|s,0^+)u_y^{\rm scat,+}(s)=&~G(\ner|s,0^+)f^\inc_+(s)+\\
&B(s)\lf( G(\ner|s,0^+)u^{\rm scat,-}(s)-G(\ner|s,0^-)u^{\rm scat,+}(s)\rg)\nonumber
\end{align}
Finally, from the identities above, and adding~\eqref{eq:above} and~\eqref{eq:below}, the integral representation formula~\eqref{eq:exact_rep_formula} for the field $u^{\rm scat}$ solution of the  boundary value problem~\eqref{eq:huygens_scattering} is obtained.

\section{Sommerfeld-integral Green's function approximation}\label{sec:Somm_int}
This appendix in devoted to the derivation of a Sommerfeld-integral~\cite{chew1995waves} representation of the proto-Green's function $G_{\rm p}$ used in the SIE derivations presented in Section~\ref{eq:Somm_int} above.  

As was mentioned above in Section~\ref{eq:Somm_int},  $G_{\rm p}$ satisfies  both the Helmholtz equation~\eqref{eq:Helm_PS} and the radiation condition, but instead of the transition conditions~\eqref{eq:BC_alpha}-\eqref{eq:BC_beta}, it satisfies the locally uniform transition conditions~\eqref{eq:approx_Green}. In order to find an expression for $G_{\rm p}$ we first note that since both metasurface parameters $\alpha$ and $\beta$ are taken to be functions of the source point $\ner'=(x',y')$, they are constant as functions of $\ner=(x,y)$ and thus it is possible to find an exact expression for $G_{\rm p}$ by standard Fourier transform techniques. In fact, for a source point $\ner'\in\Omega_+$ the proto-Green's function $G_{\rm p}$ can be interpreted as the total field produced by the incident field
\begin{equation}\label{eq:point_source_integral}G^\inc(\ner|\ner') := \frac{i}{4}H_0^{(1)}(k|\ner-\ner'|)= \frac{i}{4\pi}\int\displaylimits_{-\infty}^\infty \frac{\e^{ik_x(x-x')+ik_y|y-y'|}}{k_y}\de k_x. \end{equation}
 The square root  $k_y=k_y(k_x):=\sqrt{k^2-k_x^2}$ is defined  in the complex $k_x$~plane as the product  $\sqrt{k-k_x}\sqrt{k+k_x}$ where the first square root has a branch cut along the positive imaginary axis, and the second one has a branch cut along the negative imaginary axis. Fig.~\ref{fig:k_y} depicts the domain of definition of $k_y$ along with the curves in the complex plane where the real and imaginary parts of $k_y$ change sign.
  
  Using $G^\inc$ as incident field, the total field---which corresponds to $G_{\rm p}$---can be expressed as
\begin{subequations}\begin{equation}
 G_{\rm p}(\ner|\ner')= \lf\{\begin{array}{ccc} 
G^\inc(\ner|\ner') + G^r(\ner|\ner'), &\ner\in\Omega_+,\ner'\in\Omega_+,\medskip\\
G^t(\ner|\ner'),&\ner\in\Omega_-,\ner'\in\Omega_+,
\end{array}\rg.
\end{equation}
where the reflected and transmitted fields admit the integral representations 
\begin{eqnarray}
G^r (\ner|\ner')&:=&\frac{i}{4\pi}\circint\displaylimits_{-\infty}^\infty R\lf(k_x,x'\rg)\frac{\e^{i k_x (x-x')+ik_y |y+y'|}}{k_y}\de k_x,\label{eq:ref_GF}\\
G^t (\ner|\ner')&:=&\frac{i}{4\pi}\circint\displaylimits_{-\infty}^\infty T\lf(k_x,x'\rg)\frac{\e^{i k_x (x-x')+ik_y |y-y'|}}{k_y}\de k_x, 
\end{eqnarray}\label{eq:G_up}\end{subequations}  in terms of the reflection and transmission coefficient $R$ and $T$ defined in~\eqref{eq:ref_coeff_approx} and~\eqref{eq:trans_coeff_approx}, respectively. The special integral sign ``$\circint$" introduced in~\eqref{eq:G_up} refers to the fact that the path of integration passes below (resp. above) any pole that the integrands may have on the positive (resp. negative) real $k_x$-axis. For the sake of definiteness in what follows of this paper the integral sign $\circint$ refers to a contour integral along the path $SC$ depicted in Fig.~\ref{fig:steepest}.

In order to establish the validity of~\eqref{eq:G_up}, we note that for~$G_{\rm p}$ to satisfy the locally uniform transition conditions~\eqref{eq:approx_Green}, $R$ and $T$ have to related by the equations
\begin{equation}\label{eq:system_RT}\begin{split}
 ik_y \lf(1-R-T\rg)=&\ k\alpha(1+R+T),\\
ik_y\lf(1-R+T\rg)=&\  k\beta\lf(1+R-T\rg),\end{split}\end{equation}
where for notational simplicity we have let $R=R(k_x,x')$, $T=T(k_x,x')$, $\alpha=\alpha(x')$ and $\beta=\beta(x')$. Solving for $R$ and $T$ from~\eqref{eq:system_RT} we obtain the expressions in~\eqref{eq:ref_coeff_approx} and~\eqref{eq:trans_coeff_approx} utilized in the previous section. 

Similarly,  {it follows from the symmetric of $G_{\rm p}$ established in Appendix~\ref{app:symmetry}} that for a point source $\ner'\in\Omega_-$ the total field $G_{\rm p}$ takes the form 
\begin{equation}
 G_{\rm p}(\ner|\ner') = \lf\{\begin{array}{ccc} 
G^t(\ner|\ner'),&&\ner\in\Omega_+,\ner'\in\Omega_-,\medskip\\
G^\inc(\ner|\ner') + G^r(\ner|\ner') ,&&\ner\in\Omega_-,\ner'\in\Omega_-.
\end{array}\rg.\label{eq:G_dw}
\end{equation}
 
\section{Integral-equation formulation for corrections\label{sec:integral_eq_corrections}}
This appendix is devoted to the derivations of the SIEs~\eqref{eq:system_abstract}. To simplify the notation, we first define the functions $g_i^e(s,\sigma)= G_{\rm p}(s,0^e|\sigma,0^i)$ and $\p_yg_i^e(s,\sigma)= G_{\rm p,y}(s,0^e|\sigma,0^i)$ for~$s,\sigma\in\R$, where the indices $i$ and $e$ correspond to the symbols ``+" or ``-" that refer to the limit values (from above and below, respectively) on $\Gamma$.  

From the Sommerfeld-integral representation of $G_{\rm p}$ in~\eqref{eq:G_up} and~\eqref{eq:G_dw} it thus follows that
\begin{subequations}\begin{equation}
g_i^e(s,\sigma) = \frac{i}{4\pi}\circint\displaylimits_{-\infty}^\infty\hat g_i^e(k_x,\sigma)\e^{ik_x (s-\sigma)}\de k_x\ \mbox{and}\  \p_yg_i^e(s,\sigma) = \frac{i}{4\pi}\circint\displaylimits_{-\infty}^\infty \p_y\hat g_i^e(k_x,\sigma)\e^{ik_x (s-\sigma)}\de k_x,
\end{equation}
where
\begin{equation}\begin{split}
\hat g_+^+(k_x,\sigma)=&~\hat g_-^-(k_x,\sigma)=\frac{1}{k_y}\{1+R(k_x,\sigma)\} = \frac{1}{k_y}\lf\{2-\widehat\Gamma_\alpha(k_x,\sigma)-\widehat\Gamma_\beta(k_x,\sigma)\rg\},\\
\hat g_+^-(k_x,\sigma)=&~\hat g_-^+(k_x,\sigma)=\frac{1}{k_y}T(k_x,\sigma)=\frac{1}{k_y}\lf\{-\widehat\Gamma_\alpha(k_x,\sigma)+\widehat\Gamma_\beta(k_x,\sigma)\rg\},\\
\p_{y}\hat g_+^+(k_x,\sigma)=&-\p_{y}\hat g_-^-(k_x,\sigma)=i\lf\{1+R(k_x,\sigma)\rg\}=2i-i\lf\{\widehat\Gamma_\alpha(k_x,\sigma)+\widehat\Gamma_\beta(k_x,\sigma)\rg\}\\
\p_{y}\hat g_+^-(k_x,\sigma)=&-\p_{y}\hat g_-^-(k_x,\sigma)=-iT(k_x,\sigma)=i\lf\{\widehat\Gamma_\alpha(k_x,\sigma)-\widehat\Gamma_\beta(k_x,\sigma)\rg\},
\end{split}\label{eq:FT_kerns}\end{equation}\label{eq:FT_kerns}\end{subequations}
with  $\widehat\Gamma_\alpha$ and $\widehat\Gamma_\beta$  being defined in~\eqref{eq:terms_RT}.  

 Next we introduce the boundary integral operators
 \begin{equation}
\lf(S_i^{e}\phi\rg)(s) = \int\displaylimits_{-\infty}^\infty g_i^e(s,\sigma)\phi(\sigma)\de \sigma\mbox{ and } \lf(\p_yS_i^{e}\phi\rg)(s) = \int\displaylimits_{-\infty}^\infty \p_yg_i^e(s,\sigma)\phi(\sigma)\de \sigma,\quad s\in\R,
\label{eq:int_ops}\end{equation} (that must be interpreted in the sense of distributions) which   arise when taking  limits of~\eqref{eq:int_rep} and its normal derivative on $\Gamma$. From the integral representation~\eqref{eq:int_rep}  it follows that
\begin{equation}\begin{array}{lllllll} 
\ljmp u^{\rm scat}\rjmp(s)= (S_+^+-S_+^-)(\varphi+\psi)(s),&\quad
\slavg u^{\rm scat}\sravg(s)= (S_+^++S_+^-)(\varphi-\psi)(s),\medskip\\
\ljmp u^{\rm scat}_y\rjmp(s)= (\p_yS_+^+-\p_yS_+^-)(\varphi-\psi)(s),&\quad
\lavg u^{\rm scat}_y\ravg(s)= (\p_yS_+^++\p_yS_+^-)(\varphi+\psi)(s),
\end{array}\label{eq:jmp_avg}\end{equation}
in terms of the integral operators~\eqref{eq:int_ops}, where we have utilized the identities  $S_+^+=S_-^-$, $S_+^-=S_-^+$, $\p_yS_+^+=-\p_yS_-^-$ and $\p_yS_+^-=-\p_y S_-^+$ that result directly from~\eqref{eq:FT_kerns}.  The  uncoupled system of SIEs
\begin{subequations}\begin{eqnarray}
 \{\p_yS_+^+-\p_yS_+^-+ik\alpha(s)(S_+^++S_+^-)\}\mu_1(s) &=& -2ik\alpha(s) u^\inc(s,0) \\
\{\p_yS_+^++\p_yS_+^-+ik\beta(s)(S_+^+-S_+^-)\}\mu_2(s) &=&-2u^\inc_y(s,0)
\end{eqnarray}\label{eq:IEs}\end{subequations}
for the new  density functions $\mu_1=\varphi-\psi$ and $\mu_2 = \varphi+\psi$ is thus obtained by substituting~\eqref{eq:jmp_avg} in the transition conditions~\eqref{eq:bc_1}-\eqref{eq:bc_2}.

To show that SIE~system~\eqref{eq:IEs} is in fact  of the second-kind, we need to further study the properties of the integral operators on the left-hand-side of~\eqref{eq:IEs}. Such properties can be determined from the regularity of the integral kernels~$g_i^e$ and $\p_yg_i^e$ which can in turn  be derived from decay estimates for their Fourier transforms in~\eqref{eq:FT_kerns}. Consequently, utilizing the properties of the Fourier transform, it can be shown that the integral kernels in~\eqref{eq:IEs} can be expressed as 
\begin{equation}\begin{split}
 \p_yg_+^+(s,\sigma)-\p_yg_+^-(s,\sigma)+ik\alpha(s)(g_+^+(s,\sigma)+g_+^-(s,\sigma)) =&~ -\delta_s+K_1(s,\sigma),\\
  \p_yg_+^+(s,\sigma)+\p_yg_+^-(s,\sigma)+ik\beta(s)(g_+^+(s,\sigma)-g_+^-(s,\sigma)) =&~ -\delta_s+K_2(s,\sigma), 
\end{split}\end{equation}
where $\delta_s$ denotes the Dirac delta distribution supported at $s$ and 
\begin{subequations} \begin{eqnarray}
 K_1(s,\sigma) &:=&\frac{k}{2\pi}\lf\{\alpha(\sigma)-\alpha(s)\rg\} \circint\displaylimits_{-\infty}^\infty\frac{\e^{ik_x(s-\sigma)}}{k_y+k\alpha(\sigma)}\de k_x,\\  K_2(s,\sigma) &:=&\frac{k}{2\pi} \lf\{\beta(\sigma)-\beta(s)\rg\}\circint\displaylimits_{-\infty}^\infty\frac{\e^{ik_x(s-\sigma)}}{k_y+k\beta(\sigma)}\de k_x.
\end{eqnarray}\label{eq:int_ker_IE}\end{subequations}
Using the properties of the Dirac delta distribution and defining the integral operators  
\begin{equation}\label{eq:IOps}
\operatorname T_j[\mu](s) := \int\displaylimits_{-\infty}^\infty K_j(s,\sigma) \mu(\sigma)\de\sigma,\quad s\in\R,\ j=1,2,
\end{equation} we finally conclude that~\eqref{eq:IEs} can be equivalently expressed, in abstract form, as the SIEs~\eqref{eq:system_abstract} for the $\mu_j$, $j=1,2$.  

The key fact about $\operatorname T_j$ for its use in our series of corrections is that $\operatorname T_j \to 0$ as $\varepsilon \to 0$.   We demonstrate this numerically in Appendix~\ref{sec:conv_neu_ser}.  Analytically it occurs because the coefficients $\alpha(\sigma)-\alpha(s)$ and $\beta(\sigma)-\beta(s)$ vanish as $\varepsilon\to0$ for continuous functions $\alpha(x)=a(\varepsilon x)$ and $\beta(x)=b(\varepsilon x)$.  But, as discussed in Section~\ref{sec:conclusions}, a technically challenging asymptotic analysis is required to rigorously demarcate the function spaces and norms for which $\operatorname T_j \to 0$ and to determine the rate of convergence, which we relegate to future work.

{\section{Generalized laws of reflection and transmission\label{app:GLRT}}
This appendix is devoted to the derivation of the so-called generalized laws of reflection and transmission~\cite{Yu:2011eya,Yu:2014hq}. As it turns out, these laws can be derived from our zeroth order approximation. To see this, consider metasurface parameters of the form
\begin{equation}
\alpha(s)=\alpha_0 + \tilde \alpha(s)\quad\mbox{and}\quad \beta(s)=\beta_0+\tilde\beta(s),\quad s\in\R,
\label{eq:const_vary}\end{equation}
where $\alpha_0$ and $\beta_0$ are constants and $\tilde\alpha$ and $\tilde\beta$ are bounded functions. 
The total field resulting from the scattering of a planewave $u^\inc(\ner) = \e^{i\bold{k}^{\inc}\cdot\ner}$, $\bold k^\inc = k(\cos\theta^\inc,\sin\theta^\inc)$, $-\pi/2\leq\theta^\inc\leq 0$,  off of the metasurface can be expressed as 
$ u=v+u_0$ where 
$$
u_0(\ner) = \lf\{\begin{array}{ccc} \e^{i{\bold k}^\inc \cdot\ner}+R_0(\theta^\inc)\e^{i\bold k^r\cdot\ner},&\ner\in\Omega_+,\smallskip\\
T_0(\theta^\inc)\e^{i\bold k^\inc\cdot \ner},&\ner\in\Omega_-,\end{array}\rg.
$$
with $\bold k^r = k(\cos\theta^\inc,-\sin\theta^\inc)$ and  
$$
R_0(\theta) := 1-\frac{\alpha_0}{|\sin\theta|+\alpha_0}-\frac{\beta_0}{|\sin\theta|+\beta_0}\andtext T_0(\theta) := -\frac{\alpha_0}{|\sin\theta|+\alpha_0}+\frac{\beta_0}{|\sin\theta|+\beta_0},
$$ is the total field resulting from the scattering of the planewave $u^\inc$ off of a metasurface with constant interface parameters $\alpha_0$ and $\beta_0$. 
The field $v=u-u_0$, on the other hand, satisfies the Helmholtz equation in $\R^2\setminus\Gamma$, the radiation condition, and the transition conditions
\begin{equation*}\begin{split}
\ljmp v_y\rjmp= -ik\alpha\lavg v\ravg -ik\tilde\alpha\lavg u_0\ravg \mbox{ and }
\lavg v_y\ravg= -ik\beta\ljmp v\rjmp -ik\tilde\beta\ljmp u_0\rjmp \quad\mbox{on}\quad\Gamma.
\end{split}\end{equation*} 

Letting then $G_0=G_{\rm p}$ denote the Green's function in~\eqref{eq:G_up}~and~\eqref{eq:G_dw} corresponding to constant interface parameters $\alpha_0$ and $\beta_0$, we obtain from the discussion in Section~4 that the zeroth-order approximation (i.e.,~\eqref{eq:int_rep_ord} with $N=0$) of~$v$ is given by 
\begin{equation}
 v_0(\ner)= \int\displaylimits_{-\infty}^\infty\lf\{G_0(\ner|\sigma,0^+)\varphi_0(\sigma)- G_0(\ner|\sigma,0^-)\psi_0(\sigma)\rg\}\de \sigma,\quad \ner\in \R^{2}\setminus\Gamma, 
\label{eq:approx_GLRT}\end{equation}
where letting $$
\tilde A(s) = ik\lf\{\frac{\tilde\alpha(s)+\tilde\beta(s)}{2}\rg\}\mbox{ and } \tilde B(s) = ik\lf\{\frac{\tilde\alpha(s)-\tilde\beta(s)}{2}\rg\},
$$
 the approximate densities $\varphi_0$ and $\psi_0$ are given by
\begin{equation}\begin{split}
\varphi_0(s) =&-\lf\{\tilde A(s) T_0(\theta^\inc) +\tilde B(s) (1+R_0(\theta^\inc))\rg\}\e^{iks\cos\theta^\inc},\\
\psi_0(s) =&~\lf\{\tilde B(s) T_0(\theta^\inc)+\tilde A(s) (1+R_0(\theta^\inc))\rg\}\e^{iks\cos\theta^\inc}.
\end{split}\end{equation}

Replacing $G_0$ in~\eqref{eq:approx_GLRT} by its far-field approximation, derived in Appendix~E, we find that
\begin{equation}\label{eq:far_field_v0}
 v_0(\ner)\sim \frac{\e^{ik|\ner|}}{\sqrt{|\ner|}}  v_\infty(\theta,\theta^\inc)\quad\mbox{as}\quad |\ner|\to\infty,
\end{equation}
where, after some algebraic manipulations, the far-field pattern $ v_\infty(\theta|\theta^\inc)$ can be expressed as
\begin{subequations}\label{eq:patterns}\begin{equation}\begin{split}
 v_\infty(\theta,\theta^\inc)=&~ \sqrt{\frac{2k}{\pi}} \e^{i\frac{3\pi}{4}} |\sin(\theta^\inc)\sin(\theta)|
 \lf\{\begin{array}{llll}
\displaystyle v^\alpha_\infty(\theta,\theta^\inc)+v^\beta_\infty(\theta,\theta^\inc),& \theta\in(0,\pi),\medskip\\
\displaystyle v^\alpha_\infty(\theta,\theta^\inc)-v^\beta_\infty(\theta,\theta^\inc),&\theta\in(-\pi,0),
\end{array}\rg.\end{split}\end{equation}
where
\begin{equation}\begin{split}
v_\infty^\alpha(\theta,\theta^\inc) =&\int_{-\infty}^\infty\frac{\tilde\alpha(\sigma)\e^{ik\sigma(\cos\theta^\inc-\cos\theta)}}{(|\sin\theta|+\alpha_0)(|\sin\theta^\inc|+\alpha_0)}\de \sigma\ \mbox{ and }\\ v_\infty^\beta(\theta,\theta^\inc)=&\int_{-\infty}^\infty\frac{\tilde\beta(\sigma)\e^{ik\sigma(\cos\theta^\inc-\cos\theta)}}{(|\sin\theta|+\beta_0)(|\sin\theta^\inc|+\beta_0)}\de \sigma.
\end{split}\end{equation}\label{eq:far_field_pattern}\end{subequations}

From~\eqref{eq:patterns} it thus follows that the far-field pattern $v_\infty$ in~\eqref{eq:far_field_v0} would correspond to a linear combination of planewaves with wavevectors $\bold k_\alpha=k(\cos\theta_\alpha,\sin\theta_\alpha)$ and $\bold k_\beta = k(\cos\theta_\beta,\sin\theta_\beta)$ if  $v^\alpha_\infty$ and  $v^\beta_\infty$ were Dirac delta distributions supported at angles $\theta_\alpha$ and $\theta_\beta$, respectively.   Formally, this can be achieved by selecting $$\tilde \alpha(s) = c_\alpha\e^{ikd_\alpha s} \andtext\tilde \beta(s) = c_\beta\e^{ikd_\beta s},$$ with $c_\alpha,c_\beta\in\C$ and $d_\alpha$ and $d_\beta$ being such  that
$$
d_\alpha = \cos\theta_\alpha - \cos\theta^\inc  \quad\mbox{and}\quad  d_\beta =\cos\theta_\beta-\cos\theta^\inc.
$$

Finally, the expressions in~\eqref{eq:GLRT} are obtained by letting $\alpha_0=\beta_0=-\sin\theta^\inc,$ $c_\alpha=\pm\alpha_0$, and $c_\beta=\mp\beta_0$, for which a minimal reflection off of the metasurface is achieved.}

\section{On the convergence of the Neumann series~\eqref{eq:neu_ser}\label{sec:conv_neu_ser}}

\begin{figure}[h!]
\centering	
\includegraphics[scale=0.45]{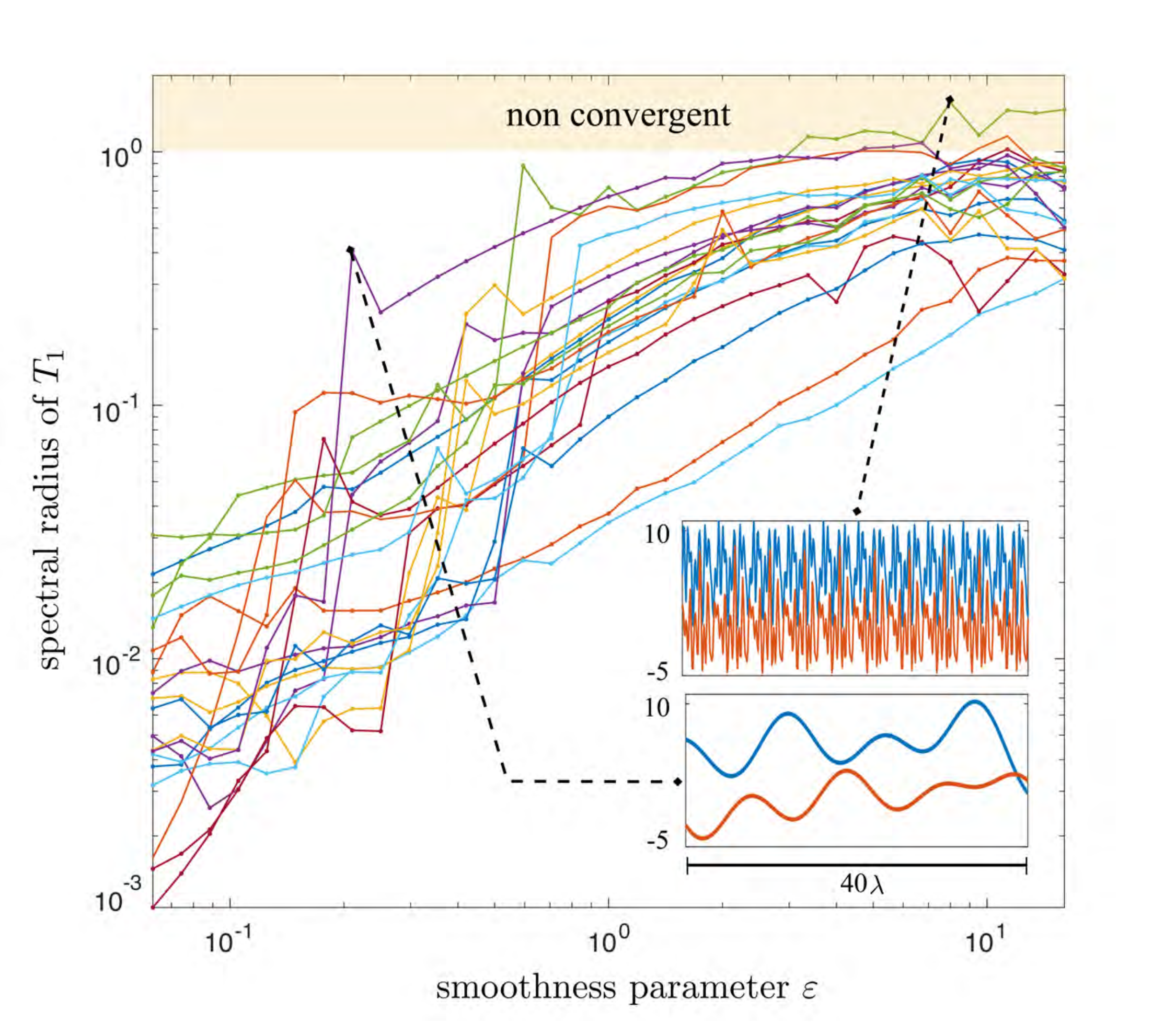}\\
\caption{Spectral radius of the discretized integral operator $T_1$ as a function of the smoothness of the interface parameter $\alpha$ in~\eqref{eq:rand_alpha}. The yellow painted strip indicates the range of parameters $\varepsilon$ for which the Neumann series diverges.}\label{fig:ex_4}
\end{figure}

 In our next appendix we consider an example to study the convergence of the Neumann series approximation~\eqref{eq:trunc_Neu_series} by examining the dependence of the spectral radii of the discretized integral operators $\operatorname T_j$, $j=1,2$, on the smoothness of the metasurface parameters $\alpha$ and $\beta$. 
We consider here the discretized version of the SIEs~\eqref{eq:system_abstract} which take the form $( I- T_j)\boldsymbol \mu_j =\boldsymbol g_j$, where $ I$ is the identity matrix, $ T_j$ is the discretized integral operator by the method described in Appendix~\ref{sec:numerics}, $\boldsymbol\mu_j$ is the unknown vector, and $\boldsymbol g_j$ is the discretized interface data. It is easy to show that if $\| T_j\|<1$ in some matrix norm, then the relative error in the discretized Neumann series approximation can be bounded by 
\begin{equation}
\frac{||\boldsymbol \mu_j-\boldsymbol \mu^{(N)}_j||}{\|\boldsymbol \mu_j\|}\leq \| T^{N+1}_j\|\leq c[\rho( T_j)]^{N+1}
\end{equation}
 for some constant $c>0$ where  $\rho( T_j) = \max_n|\lambda_n(  T_j)|$ denotes the spectral radius of $ T_j$. This bound shows that the spectral radius provides an approximate rate of convergence of the Neumann series approximation as $N$ increases. Moreover, it can be shown that the Neumann series for the discrete linear system converges if and only if the $\rho( T_j)<1$. In the following example then, we consider a metasurface parameter $\alpha(x) = a(\varepsilon x)$ given by the truncated Fourier series
\begin{equation}\label{eq:rand_alpha}
\alpha(x) =a(\varepsilon x):=\sum_{\ell=0}^{10} c_\ell\e^{i\ell \varepsilon x},\quad x\in\R,
\end{equation}
where the coefficients $c_\ell$ are randomly generated from a uniform distribution and are also adjusted so that the constrain  $\real\alpha\geq 0$ is satisfied. Clearly, $\varepsilon>0$ is a parameter that controls the smoothness of $\alpha$. Fig.~\ref{fig:ex_4} displays the spectral radius of the matrix $T_1$ in log-log scale for a range of values of~$\varepsilon$. There results demonstrate that  convergence of the Neumann series is in fact  expected for a large range of values of $\varepsilon$, including some of those that give rise to quite rough interface parameters $\alpha$.

\section{The far field of $G_{\rm p}$  and its relationship  with $G_{\rm p}^{\rm ff}$\label{sec:far_field}}
This Appendix presents a detailed asymptotic analysis that establishes rigorously the relationship between $G_{\rm p}$ and $G_{\rm p}^{\rm ff}$, as well as the existence of guided modes.

To establish the relationship between $G_{\rm p}^{\rm ff}$ and $G_{\rm p}$, we derive the far-field asymptotic approximation (as $|\ner|\to\infty$) of the proto-Green's function $G_{\rm p}$ given in~\eqref{eq:G_up} and~\eqref{eq:G_dw}. 
In order to do so we resort to the method steepest descents for which we follow the analysis of Sommerfeld integrals presented~\cite[Chapter 8]{bleistein2012mathematical}. Similar saddle point calculations can also be found in classical references on layered media scattering, such as~\cite{brekhovskikh2012waves,chew1995waves}.

Assuming first that $\ner'\in\Omega_+$ and letting~$\ner = |\ner|(\cos\theta,\sin\theta)$, $\theta\in(-\pi,0)\cup(0,\pi)$, we have that the resulting reflected and transmitted fields in~\eqref{eq:G_up} can be expressed as
 \begin{eqnarray}
G^r(\ner|\ner')= \circint_{\mathit{SC}} q_r\lf(k_x,\ner'\rg)\e^{|\ner|\phi(k_x)}\de k_x\andtext G^t(\ner|\ner')= \circint_{\mathit{SC}} q_t\lf(k_x,\ner'\rg)\e^{|\ner|\phi(k_x)}\de k_x\label{eq:asymp_int_R}
 \end{eqnarray}
in terms of the phase  and amplitude functions defined as
\begin{eqnarray}
\phi(k_x) &:= & ik_x\cos\theta+ik_y|\sin\theta|,\label{eq:phase_function}\\
q_r\lf(k_x,\ner'\rg) &:=&\frac{i R(x',k_x)}{4\pi}\frac{\e^{-ik_x x'+ik_yy'}}{k_y},\label{eq:amplitude_func}\\
q_t\lf(k_x,\ner'\rg) &:=&\frac{i T(x',k_x)}{4\pi}\frac{\e^{-ik_x x'+ik_yy'}}{k_y},\label{eq:amplitude_func_t}
\end{eqnarray}
respectively.  Note that in this case ($\ner'\in\Omega_+$) we are interested in $G^r$ for $\theta\in(0,\pi)$ and in $G^t$ for $\theta\in(-\pi,0)$.

Three kinds of critical points have to be taken in account in the steepest descent method approximation of the integrals~\eqref{eq:asymp_int_R}, namely,  saddle points of the phase function $\phi$,  (possible) poles singularities of the integrands $q_r$ and $q_t$, and the branch points of the square root  $k_y=k_y(k_x)=\sqrt{k^2-k_x^2}$.   

\begin{figure}[h!]
\centering	
\includegraphics[scale=0.82]{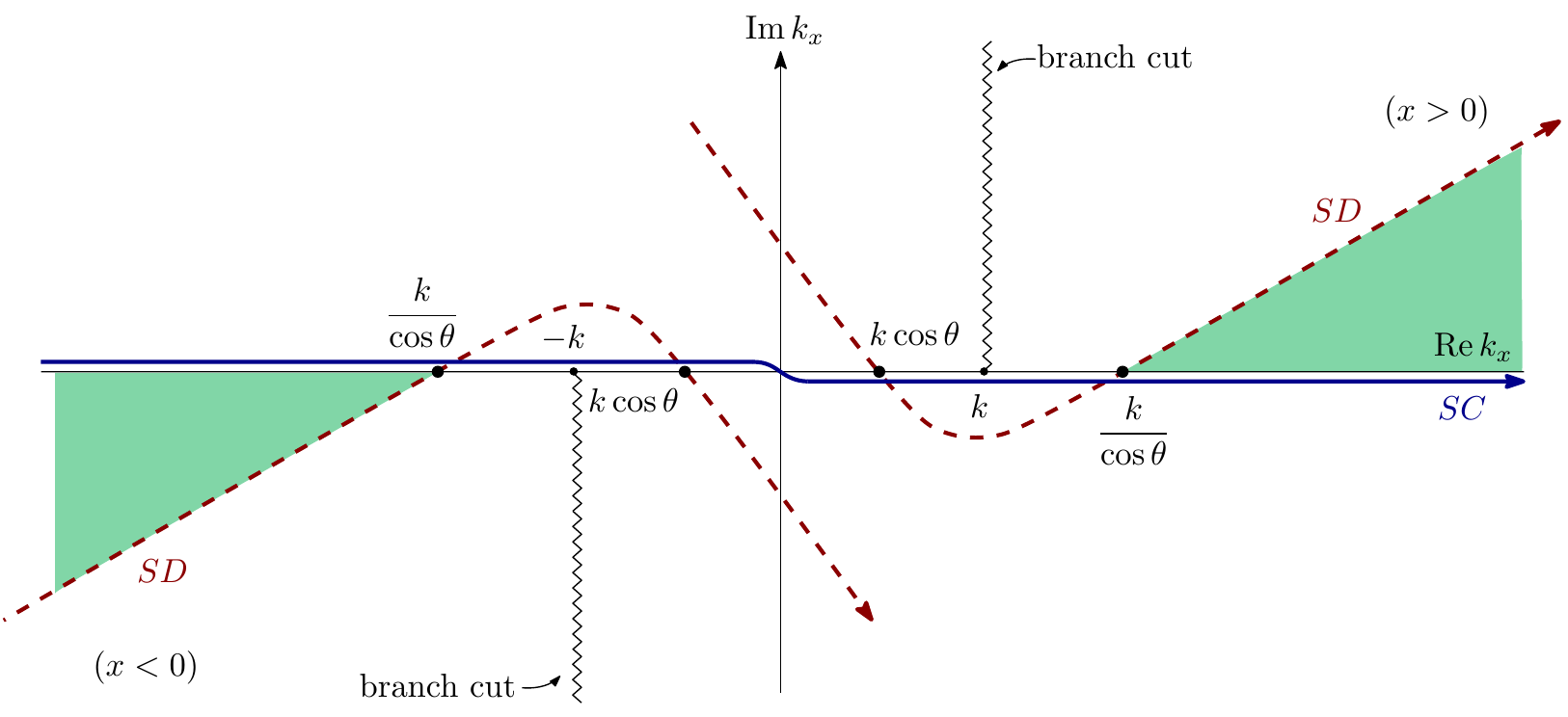}
\caption{Sommerfeld contour $\mathit{SC}$ (continuous  blue line) and steepest descent contour $\mathit{SD}$ (dashed red line) utilized in the evaluation of the integral~\eqref{eq:asymp_int_R}. Poles of the $\widehat \Gamma_\alpha$ and $\widehat \Gamma_\beta$ lying in the shaded regions contribute to the far-field asymptotic expansion of the approximate Green's function.}\label{fig:steepest}
\end{figure}

We first  consider the saddle points of $\phi$, which correspond to solutions $k^*_x\in\C$ of the algebraic equation $\phi'(k^*_x)=0$. In view of 
$\phi'(k_x) = i\cos\theta-ik_x|\sin\theta|/k_y$ follows that there  is only one saddle point on $SC$ given by $k_x^* =k \cos\theta = kx/|\ner|$ at which $
\phi(k_x^*) = ik$.  The steepest descent directions from $k_x^*$, on the other hand, are given by the angles $3\pi/4$ and $-\pi/4$, which were obtained from $\phi''(k_x^*) = -i/(k\sin^2\theta)\neq 0$. We then conclude that the steepest descent path $SD$ is given implicitly by the equation $\imag \phi(k_x)=k$ from which it can be shown that $\mathit{SD}$ intersects $\mathit{SC}$ again at $k_x=k\sec\theta$ and that 
\begin{subequations}\begin{equation}
\imag k_x = \real k_x |\cot\theta|-k|\csc \theta|\quad \mbox{as}\quad |k_x|\to\infty\quad\mbox{if}\quad x>0,
\end{equation}
and
\begin{equation}
\imag k_x = \real k_x|\cot\theta|+k|\csc\theta|\quad \mbox{as}\quad |k_x|\to\infty\quad\mbox{if}\quad x<0.
\end{equation}\label{eq:asymp_SD}\end{subequations}
 Fig.~\ref{fig:steepest} depicts the steepest descent paths for $x>0$ and $x<0$. $\mathit{SD}$ coincides with $\mathit{SC}$ when $x=0$.

 \begin{figure}[h!]
\centering	
\includegraphics[scale=0.85]{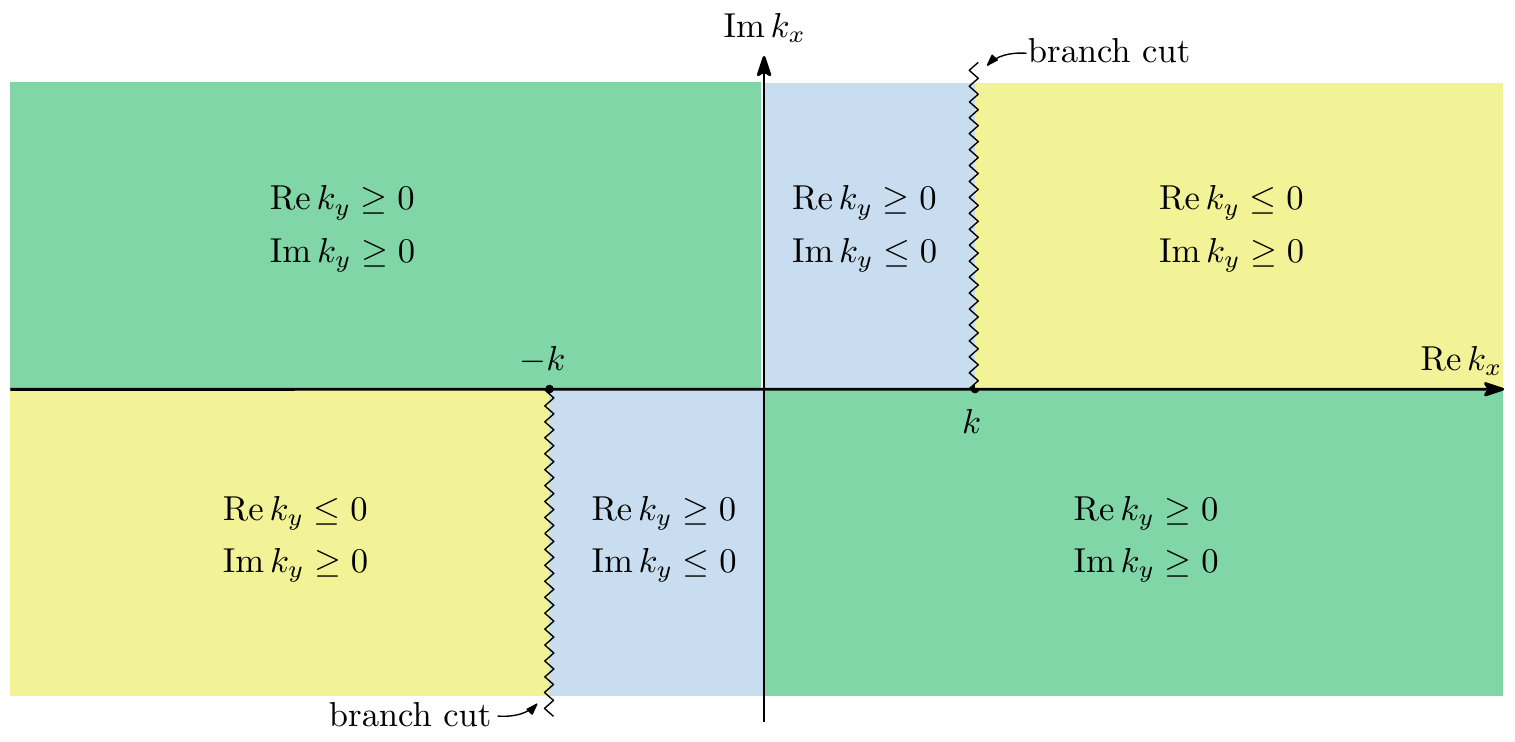}
\caption{Signs of the real and imaginary parts of square root  $k_y(k_x)=\sqrt{k_x^2-k^2}$ in the complex $k_x$~plane.}\label{fig:k_y}
\end{figure}

With this information in hand we then proceed to deform the Sommerfeld contour $\mathit{SC}$ to the steepest descent contour $\mathit{SD}$ that passes through the saddle point $k_x^*$.  Note that $\mathit{SD}$ does not intersect the branch cuts stemming from $\pm k$ and, thus, there is no contribution to the asymptotic expansions from the branch points at $\pm k$. There might be, however, contributions arising from pole singularities  of the integrands.  In fact, from the expressions for the reflection and transmission coefficients in~\eqref{eq:ref_coeff_approx} and~\eqref{eq:trans_coeff_approx}, respectively, we have that the poles of both $q_r$~\eqref{eq:amplitude_func} and~$q_t$~\eqref{eq:amplitude_func_t}---which correspond to the poles of the functions $\widehat \Gamma_\alpha(x,x')$ and $\widehat \Gamma_\beta(x,x')$ defined in~\eqref{eq:terms_RT}---are solutions of the (independent) algebraic equations 
\begin{equation}
k_y(k_x)= -k\alpha(x')\quad\mbox{or}\quad k_y(k_x) =-k\beta(x').\label{eq:poles_GF}
\end{equation}
To find necessary and sufficient conditions on the metasurface parameters $\alpha$ and $\beta$ for such poles to exist, we first note that the conditions~$\real\alpha\geq 0$, $\real\beta\geq 0$ and  the equations~\eqref{eq:poles_GF} imply that poles of $q_r$ or $q_t$ could only exist in the regions of the complex $k_x$~plane where $\real k_y(k_x)\leq 0$. It is easy to see that such  regions amount to $D = \{\real k_x\geq k,\imag k_x\geq 0\}\cup\{\real k_x\leq -k,\imag k_x\leq 0\}$ (Fig.~\ref{fig:k_y}). Furthermore, since $\imag k_y=\sqrt{k^2-k_x^2}\geq 0$ in $D$, poles will exist if and only $\imag \alpha\leq 0$ or $\imag\beta\leq 0$. 

By the Cauchy residue theorem we thus have that the poles of $q_r$  (resp.~$q_t$), if any, will only contribute to the far-field expansion of $G^r$ (resp. $G^t$) if they lie within the region in the complex plane enclosed by $\mathit{SC}$ and  $\mathit{SD}$. In order to determine whether a pole of $q_r$ (resp. $q_t$) lies inside that region, and in view of the fact that we do not have access to an explicit parametrization of $\mathit{SD}$, we utilize the asymptotic identities~\eqref{eq:asymp_SD}. Doing so we conclude that the relevant poles---that are henceforth  denoted by $k_x^{(p)}$, $p=1,2$---must meet the conditions 
\begin{subequations}\begin{equation}
k_x^{(1)} = k\sqrt{1-\alpha^2}\ \mbox{or}\  k_x^{(2)} = k\sqrt{1-\beta^2},\ \mbox{and}\  0\leq \imag k_x^{(p)} \leq  \real k_x^{(p)}|\cot\theta|-k|\csc\theta| \ \ \mbox{if}\ \ x>0,
\end{equation}
and
\begin{equation}
k_x^{(1)} = -k\sqrt{1-\alpha^2}\ \mbox{or}\ k_x^{(2)} = -k\sqrt{1-\beta^2},\  \mbox{and}\   \real k_x^{(p)}|\cot\theta|+k|\csc\theta|\leq  \imag k_x^{(p)} \leq 0 \ \mbox{if} \ x<0.\end{equation}\label{eq:cond_poles}\end{subequations} 

Therefore,  accounting  for both saddle point and poles contributions, we obtain the following asymptotic expansions
\begin{subequations}\begin{equation}
G^r(\ner|\ner' )=\frac{\e^{ik|\ner|-ik\ner\cdot\bar\ner'/|\ner|+i\frac{\pi}{4}}}{\sqrt{8\pi k|\ner|}}R\lf(\frac{k x}{|\ner|},x'\rg) + \sum_{p=1,2} A^r_p\e^{ik_x^{(p)}|x-x'|+ik_y(k_x^{(p)})|y+y'|}+O\lf(|\ner|^{-3/2}\rg)
\label{eq:hf_final_r}
\end{equation}
and 
\begin{equation}
G^t(\ner|\ner' ) = \frac{\e^{ik|\ner|-ik\ner\cdot\ner'/|\ner|+i\frac{\pi}{4}}}{\sqrt{8\pi k|\ner|}}T\lf(\frac{kx}{|\ner|},x'\rg) + \sum_{p=1,2} A^t_p\e^{ik_x^{(p)}|x-x'|+ik_y(k_x^{(p)})|y-y'|}+O\lf(|\ner|^{-3/2}\rg)
\label{eq:hf_final_t}
\end{equation}\label{eq:hf_final}\end{subequations}
as $|\ner|\to\infty$, where $\bar\ner'$ is the image point source $\bar\ner' = (x',-y')$ (Fig.~\ref{fig:approx_GF}).   The amplitudes $A^r_1$ and $A^t_1$ of the guided waves in~\eqref{eq:hf_final} are directly obtained from  the residues of $q_t$ and $q_r$ and  read as
\begin{equation}
A^r_1=A^t_1 =   -\frac{\alpha}{2\sqrt{1-\alpha^2}} 
\end{equation}
if  a pole $k_x^{(1)}$ in~\eqref{eq:cond_poles} exists, and they equal zero otherwise. Similarly, 
\begin{equation} A^r_2 =-A^t_2 =  -\frac{\beta}{2\sqrt{1-\beta^2}}\end{equation}
if  a pole $k_x^{(2)}$ in~\eqref{eq:cond_poles} exists, and they equal zero otherwise.

 The contribution to the asymptotic expansions of the pole singularities corresponds to surface waves that travel away from the point source $\ner'=(x',y')$ and are confined to a narrow strip containing the metasurface---as they decay exponentially fast toward the upper and lower half-planes. For example, the contribution of the pole  $k_x^{(1)}=k\sqrt{1-\alpha^2}$ to the asymptotic expansion of the reflected and transmitted fields equals
\begin{equation}\label{eq:surf_wave_mode} -\frac{\alpha}{2\sqrt{1-\alpha^2}}  \e^{ik\sqrt{1-\alpha^2}|x-x'|-ik\alpha|y\pm y'|},\end{equation}
with $+$ and $-$ corresponding to the reflected and transmitted fields, respectively.

In order to establish the relationship between $G^{\rm ff}_{\rm p}$ and $G_{\rm p}$ we recall the asymptotic identity
\begin{equation}
\frac{i}{4}H_0^{(1)}(k|\ner-\ner'|) = \frac{\e^{ik|\ner|-ik\ner\cdot\ner'/|\ner|+i\frac{\pi}{4}}}{\sqrt{8\pi k|\ner|}} + O(|\ner|^{-3/2})\quad\mbox{as}\quad|\ner|\to\infty,
\label{eq:asymp_hankel}\end{equation}
that follows from $|\ner-\ner'| = \ner  - \ner\cdot\ner'/|\ner|+O(|\ner|^{-1})$ as $|\ner|\to\infty$ and the standard asymptotic expansion of the Hankel function.
 Using~\eqref{eq:asymp_hankel}, the fact that 
$(x-x')/|\ner-\ner'| = x/|\ner| + O(|\ner|^{-1})$ as $|\ner|\to\infty$,
and  comparing~\eqref{eq:KA_approx} with~\eqref{eq:G_up}, we conclude that 
\begin{equation}
 G_{\rm p}^{\rm ff}(\ner|s,0^\pm) =  G_{\rm p}(\ner|s,0^\pm)+O(|\ner|^{-3/2})\quad\mbox{as}\quad|y|\to\infty.\label{eq:relation_KA}
\end{equation}
Therefore,  $G^{\rm ff}_{\rm p}$~\eqref{eq:KA_approx} can be simply interpreted as an approximation of the proto-Green's function~$G_{\rm p}$ for point sources at the metasurface and observations points far away from the metasurface.  Note that for such configuration of source and observation points, the surface wave modes do not have any significant contribution as they decay exponentially as $|y|\to\infty$.

\begin{figure}[h!]
\centering	
\includegraphics[scale=0.45]{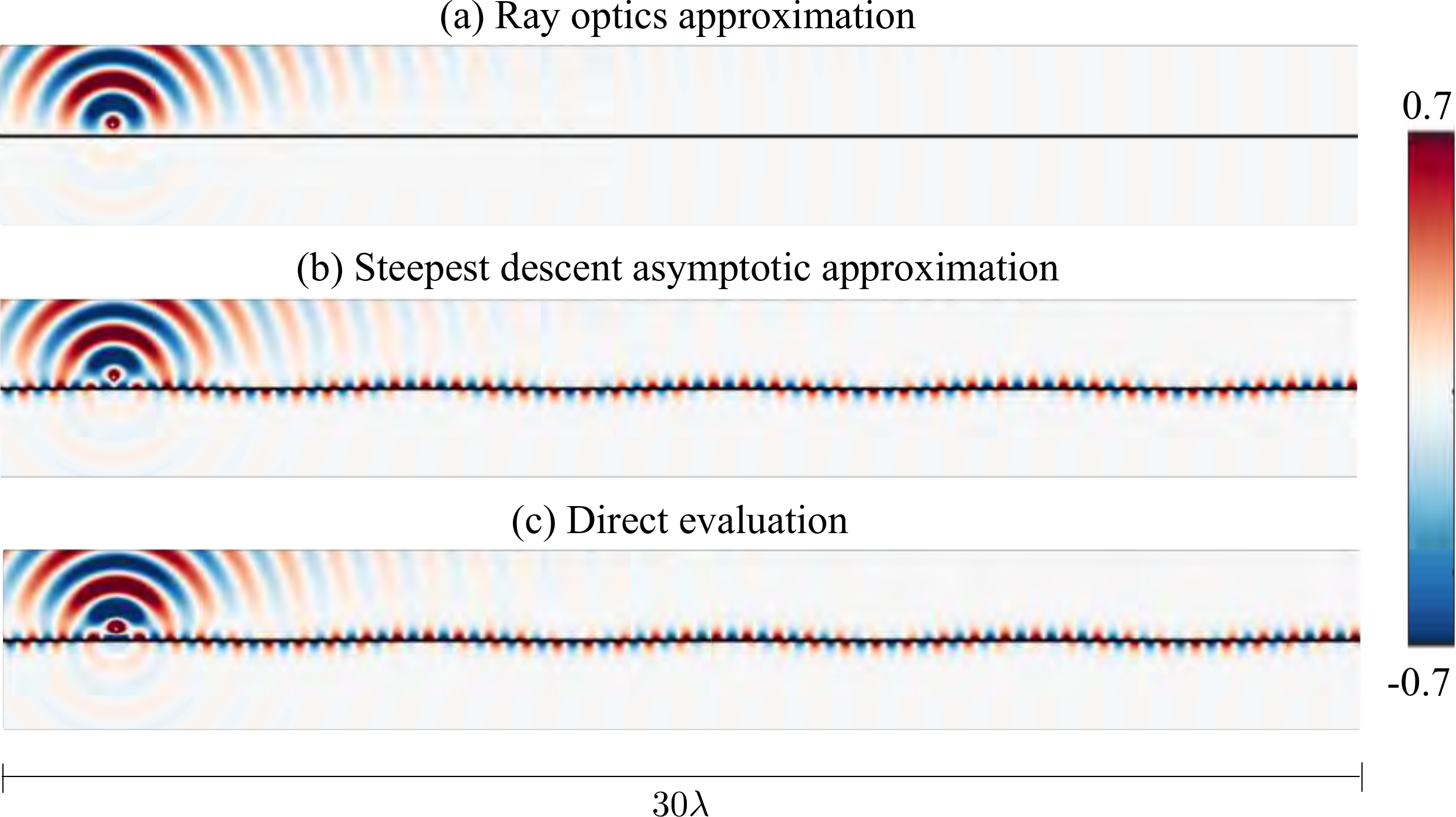}
\caption{Real part of the total field solution of the problem of scattering of the field produced by a point source by a metasurface with constant interface parameters $\alpha=-i$ and $\beta=-1.2i$. This selection of the interface parameters allow a surface wave mode of the form~\eqref{eq:surf_wave_mode} to propagate along the metasurface. The point source is located at distance $0.25\lambda$ above the metasurface. (a) $G_{\rm p}^{\rm ff}$ which in this case corresponds to the ray-optics approximation of the exact Green's function~$G$. (b) Rigorous far-field asymptotic approximation~\eqref{eq:hf_final} of $G_{\rm p}$. (c) Direct evaluation of the proto-Green's function $G_{\rm p}$ defined in~\eqref{eq:G_up} and~\eqref{eq:G_dw}.}\label{fig:ex_5}
\end{figure}

Finally, in order to demonstrate the validity of the asymptotic expansions derived in this section we present Fig.~\ref{fig:ex_5} that displays the real part of $G_{\rm p}(\ner|\ner')$ for constant interface parameters $\alpha=-i$ and $\beta=-1.2i$.
  Note that for these interface parameters the terms $\widehat \Gamma_\alpha(k_x,s)$ and $\widehat \Gamma_\beta(k_x,s)$ defined in~\eqref{eq:terms_RT} have poles on the real axis which make surface waves mode of the form~\eqref{eq:surf_wave_mode} appear in the asymptotic expansions~\eqref{eq:hf_final}. 
  
\section{Numerics}\label{sec:numerics}
In this appendix we briefly describe a high-order method for the numerical evaluation of
the Sommerfeld integrals $G^r$ and $G^t$ in~\eqref{eq:G_up}-\eqref{eq:G_dw} and the integral kernels $K_1$ and $K_2$ in~\eqref{eq:int_ker_IE}. This approach, which was originally developed for layered-media scattering problems~\cite{perez2014high} (see also~\cite[Section 2.3.5]{perez2017windowed}), is a combination of the contour-integration method described
in~\cite{Paulus:2000vr} and the the smooth-windowing approach put
forth in~\cite{MonroJr:2008te} for the evaluation of oscillatory
integrals. 

\paragraph{Proto-Green's function}\label{sec:num_SIGF}
Consider the Sommerfeld integrals $G^r$ and  $G^t$ in~\eqref{eq:G_up}-\eqref{eq:G_dw} which are given by linear combinations of integrals of the form 
\begin{equation}
\Phi(d_1,d_2) =  \frac{i}{4\pi}\circint\displaylimits_{-\infty}^\infty\frac{k\gamma}{k_y+k\gamma}\frac{\e^{i(k_x d_1+k_yd_2)/k}}{k_y}\de k_x,\quad d_1,d_2\geq 0,\label{eq:green_int}
\end{equation}
where $d_1=k|x-x'|$, $d_2 = k|y+y'|$ in the case of $G^r$ or $d_2=k|y-y'|$ in the case of $G^t$, and $\gamma=\alpha(x')$ or  $\gamma=\beta(x')$. Note that $d_j$, $j=1,2,$ are dimensionless variables. To tackle the most challenging integration scenario, in what follows we consider the case when the integrand has a pole at $k\sqrt{1-\gamma^2}$ on the real axis.  

\begin{figure}[h!]
\centering	
\includegraphics[scale=1.0]{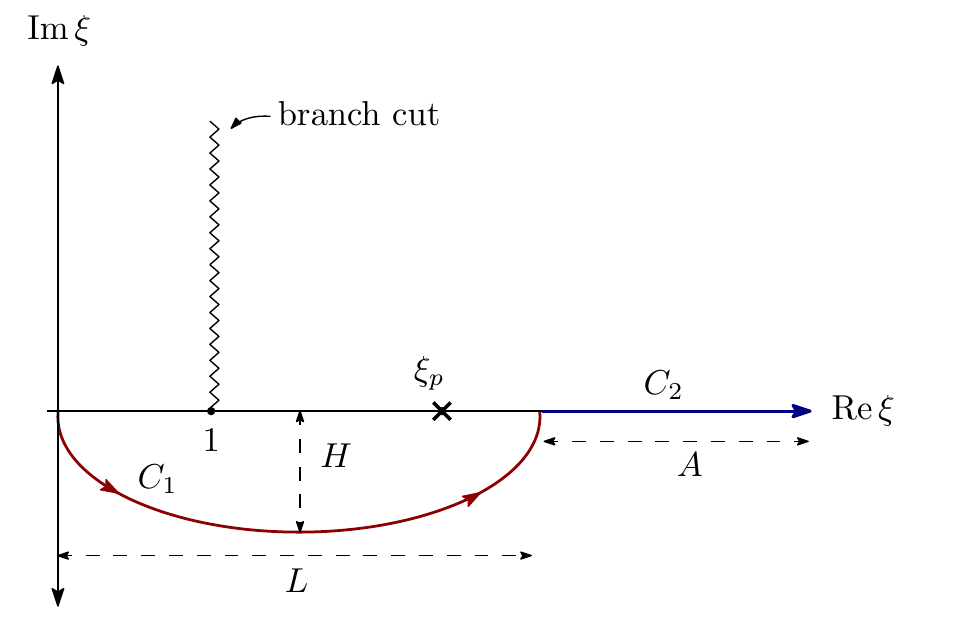}
\caption{Integration contours used in the evaluation of the proto-Green's function $G_{\rm p}$, and the integral kernels $K_1$ and $K_2$.}\label{fig:eval_path}
\end{figure}
Making use of the change of variable $\xi = k_x/k$, using the fact that $k_y(k_x)=k_y(-k_x)$, and letting 
\begin{equation}
f(\xi)= \frac{i\gamma}{\pi}\frac{\cos(\xi d_1)}{\sqrt{1-\xi^2}+\gamma}\frac{\e^{id_2\sqrt{1-\xi^2}}}{\sqrt{1-\xi^2}},
\end{equation}
we have that $\Phi$ can be expressed as $\Phi= I_1+I_2$ where $I_j=\int_{C_j} f(\xi)\de \xi$, $j=1,2,$ with  $C_1$ and $C_2$ being the contours depicted in Fig.~\ref{fig:eval_path}.  
The curve $C_1$ is a simple curve in the fourth quadrant that is parametrized by a smooth complex-valued function $\zeta:[0,\pi]\mapsto \mathbb{C}$ satisfying $\zeta(0)=0$ and $\zeta(\pi)=L:=1+\xi_p$, where $\xi_p$ is the pole of $f$ at $\sqrt{1-\gamma^2}$.  For the sake of definiteness, the curve $C_1$ is here selected as the semi-ellipse 
\begin{equation}
\zeta(t) := \left\{\frac{L(1+\cos(t+\pi))}{2}+iH\sin(t+\pi): t\in[0,\pi]\right\}
\end{equation}
that passes below all the singularities of the integrand $f$. The contour $C_2$, on the other hand, is simply the interval $[L,\infty]$ on the real axis. 

Note that on $C_1$ the function~$f(\zeta(t))$ grows exponentially as $t$
increases from 0 to $\pi/2$. Indeed,  it can be shown~\cite{perez2017windowed} that $\left|\e^{id_2\sqrt{1-\xi^2}}\right| \leq \e^{d_2H}$  and $\left|\cos(d_1\xi)\right|\leq  \e^{d_1H}$ for $\zeta\in C_1$. Thus, in order to control the exponential growth of the integrand  on $C_1$ we select $H = \lf(\max\{10,d_1+d_2\}\rg)^{-1}$. This simple procedure  ensures that the
exponential terms of $f$ remain bounded by one along $C_1$.
The resulting expression for the contour integral $I_1$ is then approximated by means of the Clenshaw--Curtis quadrature
rule~\cite{davis2007methods}---which, for the smooth integrand under
consideration, yields rapid convergence.  In view of the oscillatory
behavior of the integrand and in order to maintain
the same accuracy for all $d_1$ and $d_2$, the number of quadrature points is chosen to grow linearly with~$d_1$.

In order to evaluate the oscillatory integral $I_2$, on the other
hand, we utilize the windowing method put forth
in~\cite{MonroJr:2008te}. Using this procedure $I_2$ is approximated as
\begin{equation}
\begin{split}
I_2 \approx \int_{L}^{A+L}f(t)w_A(t-L)\de t,
\end{split}\label{eq:second_integral}
\end{equation}
where the window function $w_A$ is defined as $w_A(t) := \eta(t;cA,A)$, $A>0$, $0<c<1$,  in terms of the 
  $C^\infty(\R)$ function 
\begin{equation}
\eta(t;t_0,t_1) =\left\{
\begin{array}{cll}
1,&|t|\leq t_0,\medskip\\
\!\!\exp\left(\displaystyle\frac{2\e^{-1/u}}{u-1}\right),&\displaystyle t_0<|t|<t_1, u=\frac{|t|-t_0}{t_1-t_0},\medskip\\
0,&|t|>t_1,
\end{array}\right.\label{eq:window_function}
\end{equation}
which  equals one on $[-t_0,t_0]$ and is supported on the (bounded) interval $[-t_1,t_1]$.

 In virtue of the
oscillatory behavior of the integrand when $d_1\neq
0$, and the exponential decay of the integrand when $d_2\neq 0$, the integral on the right-hand-side of \eqref{eq:second_integral} converges
to $I_2$ faster than any negative power $\sqrt{d_1^2+d_2^2}A$ as~$A$ goes to infinity---as
proved in~\cite[Proposition~2.3.4]{perez2017windowed}.  In the special case
$d_1=d_2=0$, however, the $f$  is slowly decaying on $C_2$
and does not oscillate, thus, it leads to slow (algebraic) convergence of the
windowed-integral approximation~\eqref{eq:second_integral} to the
integral $I_2$ as $A\to\infty$. In fact, in that case the error in the windowed-integral approximation decays as 
$ O((cA)^{-1})$~\cite[Proposition~2.3.4]{perez2017windowed}. Therefore, as in the case of $I_1$, the integral on the right-hand-side of~\eqref{eq:second_integral} is here
approximated by using Clenshaw--Curtis quadrature.  The super-algebraic/exponential convergence of the windowed integral
allows $I_2$ to be approximated with a fix accuracy and a fixed
computational cost by choosing $A$ inversely proportional to~$\sqrt{d_1^2+d_2^2}$.  

\paragraph{Integral kernels}
In order to numerically evaluate the integral kernels $K_1$ and $K_2$ defined in~\eqref{eq:int_ker_IE}, we  resort to the contour-integration procedure described above. The evaluation of these kernels requires the approximation of integrals of the form 
\begin{equation}\label{eq:psi_int}
 \Psi(d) =~\circint\displaylimits_{-\infty}^\infty\frac{\e^{ik_xd/k}}{k_y+k\gamma}\de k_x,\quad d\geq 0,
 \end{equation}
where $d=k|s-\sigma|$, $\gamma=\alpha(\sigma)$ in the case of $K_1$ and $\gamma=\beta(\sigma)$ in the case of $K_2$.  Unlike~\eqref{eq:green_int}, the Fourier integral~\eqref{eq:psi_int} is only conditionally convergent for $d>0$ and diverges at $d=0$. In order to separate the singularity of $\Psi$ at $d=0$ we use identity 
\begin{equation}
H_0^{(1)}(d)=\frac{1}{\pi}\int\displaylimits_{-\infty}^\infty\frac{\e^{ik_xd/k}}{k_y}\de k_x=\frac{1}{\pi}\int\displaylimits_{-\infty}^\infty\frac{\e^{i\xi d}}{\sqrt{1-\xi^2}}\de \xi,\qquad d\geq 0,
\end{equation} which follows from~\eqref{eq:point_source_integral}, to obtain
\begin{equation}\label{eq:sing_substraction}
\Psi(d) =\pi H_0^{(1)}(d) -k\gamma \circint\displaylimits_{-\infty}^\infty\frac{\e^{ik_xd/k}}{k_y(k_y+k\gamma)}\de k_x.\end{equation}
Thus, the  integral in~\eqref{eq:sing_substraction}---that turns out to be a continuous function of $d$---can now be evaluated directly by means of the contour integration procedure presented above for all $d\geq 0$.  We thus have  $\Psi(d) = \pi H_0^{(1)}(d) -\gamma(I_1+I_2)$  where $I_j = \int_{C_j}f(\xi)\de\xi$ with $f$ now being given by $ f(\xi) = \e^{i\xi d}/[\sqrt{1-\xi^2}(\sqrt{1-\xi^2}+\gamma)]$.  Although $f$ is absolutely integrable (it decays as $\xi^{-2}$ as $|\xi|\to\infty$) a large value of $A>0$ is needed in the windowed approximation of $I_2$~\eqref{eq:second_integral} to achieve a prescribed accuracy when $d=0$. In order to improve the slow  $O((cA)^{-1})$ convergence rate as $A\to\infty$ when $d=0$, we note further that using the identity 
\begin{equation}\circint\displaylimits_{-\infty}^\infty\frac{\e^{ik_xd/k}}{k^2-k_x^2}\de k_x=-\frac{i\pi}{2k} \e^{ikd/k},\qquad d\geq 0, \end{equation}
which follows directly from Jordan's lemma and Cauchy's residue theorem,
 the integral in~\eqref{eq:sing_substraction} can be expressed  as 
\begin{equation}\label{eq:int_sub}
\circint\displaylimits_{-\infty}^\infty\frac{\e^{ik_xd/k}}{k_y(k_y+k\gamma)} \de k_x=-\frac{i\pi}{2k} \e^{ikd/k}-k\gamma\circint\displaylimits_{-\infty}^\infty\frac{\e^{ik_xd/k}}{(k^2-k_x^2)(k_y+k\gamma)}\de k_x.
\end{equation}
Therefore, since the integrand on the right-hand-side of~\eqref{eq:int_sub} decays as $\xi^{-3}$ as $|\xi|\to\infty$, the contour integration procedure described in the previous section  yields super-algebraic convergence as $A\to\infty$ for $d>0$ and now yields an $O((cA)^{-2})$ convergence rate when $d=0$.

\paragraph{Integral operators and potentials} Finally, we briefly mention that  the (improper) oscillatory integrals  in~\eqref{eq:kirchhoff_approx},~\eqref{eq:int_rep},~\eqref{eq:corrections},  and in the definition of the operators $T_1$ and $T_2$~\eqref{eq:IOps}, can be accurately truncated by means of the windowing procedure described above, in a manner similar to the windowed Green function method~\cite{bruno2016windowed,bruno2017windowed_a,bruno2017windowed}. In order to handle the logarithmic singularity of the integral kernels~\eqref{eq:IOps}, on the other hand,  standard singular integrations techniques, such as the spectrally accurate Martensen--Kussmaul quadrature rule~\cite{COLTON:2012}, can be used.

\bibliographystyle{abbrv}
\bibliography{References}

\end{document}